\documentclass[12pt]{article}

\usepackage{graphicx}
\usepackage{graphics}
\usepackage{amssymb}
\usepackage{amsmath}
\usepackage{epsf}

\usepackage[parfill]{parskip}
\usepackage[matrix,arrow]{xy}
\usepackage[usenames]{color}
\input{epsf}

\newcommand{\ba}{\begin{eqnarray*}}
\newcommand{\ea}{\end{eqnarray*}}
\newcommand{\ban}{\begin{eqnarray}}
\newcommand{\ean}{\end{eqnarray}}
\newcommand{\beq}{\begin{equation*}}
\newcommand{\eeq}{\end{equation*}}
\newcommand{\beqn}{\begin{equation}}
\newcommand{\eeqn}{\end{equation}}

\newcommand{\Tr}{{\rm Tr\,}}
\newcommand{\tr}{{\rm tr\,}}

\newcommand{\IZ}{\mathbb{Z}}
\newcommand{\IC}{\mathbb{C}}
\newcommand{\IP}{\mathbb{P}}
\newcommand{\IN}{\mathbb{N}}

\newcommand{\cO}{{\cal O}}
\newcommand{\cC}{{\cal C}}

\newcommand{\td}{\tilde}

\newcommand{\CYX}{{\mathfrak X}}

\newcommand{\Res}{\mathop{{\rm Res}\,}}
\newcommand{\Li}{\rm Li}

\newcommand{\curve}{{{\mathcal C}}}

\makeatletter
\@addtoreset{equation}{section}
\makeatother

\newcommand\encadremath[1]{\vbox{\hrule\hbox{\vrule\kern8pt
\vbox{\kern8pt \hbox{$\displaystyle #1$}\kern8pt}
\kern8pt\vrule}\hrule}}
\def\enca#1{\vbox{\hrule\hbox{
\vrule\kern8pt\vbox{\kern8pt \hbox{$\displaystyle #1$}
\kern8pt} \kern8pt\vrule}\hrule}}

\def \nn{\nonumber}

\newdimen\tableauside\tableauside=1.0ex
\newdimen\tableaurule\tableaurule=0.4pt
\newdimen\tableaustep
\def\phantomhrule#1{\hbox{\vbox to0pt{\hrule height\tableaurule width#1\vss}}}
\def\phantomvrule#1{\vbox{\hbox to0pt{\vrule width\tableaurule height#1\hss}}}
\def\sqr{\vbox{%
  \phantomhrule\tableaustep
  \hbox{\phantomvrule\tableaustep\kern\tableaustep\phantomvrule\tableaustep}%
  \hbox{\vbox{\phantomhrule\tableauside}\kern-\tableaurule}}}
\def\squares#1{\hbox{\count0=#1\noindent\loop\sqr
  \advance\count0 by-1 \ifnum\count0>0\repeat}}
\def\tableau#1{\vcenter{\offinterlineskip
  \tableaustep=\tableauside\advance\tableaustep by-\tableaurule
  \kern\normallineskip\hbox
    {\kern\normallineskip\vbox
      {\gettableau#1 0 }%
     \kern\normallineskip\kern\tableaurule}%
  \kern\normallineskip\kern\tableaurule}}
\def\gettableau#1 {\ifnum#1=0\let\next=\null\else
  \squares{#1}\let\next=\gettableau\fi\next}

\begin{document}

\begin{titlepage}
\begin{flushright}
IHES/P/10/06\\
IPHT/T10/029\\
LPTENS 10/14\\
\end{flushright}
\begin{center}
\vskip 2cm {\Huge A matrix model for the topological string I
\\ \vskip 0.2cm}
{\LARGE Deriving the matrix model}\\
\vskip 1cm {B. Eynard${}^1$, A. Kashani-Poor${}^{2,3}$, O. Marchal${}^{1,4}$}

\vskip.6cm 
{\it ${}^1$ Institut de Physique Th\'eorique,\\
CEA, IPhT, F-91191 Gif-sur-Yvette, France,\\
CNRS, URA 2306, F-91191 Gif-sur-Yvette, France.\\ \vskip0.3cm}

{\it 
$^2$ Institut des Hautes \'Etudes Scientifiques\\
Le Bois-Marie, 35, route de Chartres, 91440 Bures-sur-Yvette, France\\ \vskip0.3cm }

{\it $^3$ Laboratoire de Physique Th\'eorique de l'\'Ecole Normale Sup\'erieure, \\
24 rue Lhomond, 75231 Paris, France \\ \vskip0.3cm}

{\it ${}^4$ Centre de recherches math\'ematiques,
Universit\'e de Montr\'eal \\
C.P. 6128, Succ. centre-ville
Montr\'eal, Qu\'e, H3C 3J7, Canada.\\}

\end{center} 
\vskip 1.5cm
\begin{abstract}
We construct a matrix model that reproduces the topological string 
partition function on arbitrary toric Calabi-Yau 3-folds. This 
demonstrates,  in accord with the BKMP ``remodeling the B-model'' 
conjecture, that Gromov-Witten invariants of any toric Calabi-Yau 3-fold can be computed in terms of the spectral invariants of a spectral curve. 
Moreover, it proves that the generating function of Gromov-Witten 
invariants is a tau function for an integrable hierarchy. In a follow-up paper, we will explicitly construct the spectral curve of our 
matrix model and argue that it equals the mirror curve of the toric 
Calabi-Yau manifold.
\end{abstract}

\end{titlepage}
\newpage

\tableofcontents

\section{Introduction}

In the topological string A-model, the object of study is the moduli space of maps from a Riemann surface $\Sigma_g$ of genus $g$ to a given Calabi-Yau target space ${\mathfrak{X}}$. 
Its partition function is the generating function of Gromov-Witten invariants of ${\mathfrak{X}}$, which roughly speaking count these maps. 

In recent years, deep connections have been unrooted between the topological string on various geometries and random matrix models. A classic result in the field is that intersection numbers, which are related to the Gromov-Witten theory of a point, are computed by the Kontsevich matrix integral \cite{Kontsevich}, see also \cite{Okounkov2}. In the Dijkgraaf-Vafa conjecture \cite{DVmatrix} such a connection is obtained between the topological B-model on certain non-compact Calabi-Yau manifolds and a 1-matrix model. A novel type of matrix model \cite{marino3} inspired by Chern-Simons theory is associated to the topological string in \cite{AKMV_CS}, yielding matrix model descriptions of target spaces obtained from the cotangent space of lens spaces via geometric transition. This work is extended to chains of lens spaces and their duals in \cite{HalmagyiOkuda}.

In the 20 years that have passed since topological string theory was formulated \cite{Witten, BCOV}, various techniques have been developed for computing the corresponding partition function. The topological vertex method \cite{AKMV} solves this problem completely for toric Calabi-Yau 3-folds at large radius, furnishing the answer as a combinatorial sum over partitions. On geometries with unit first Betti number (the conifold and $\cO(-2) \rightarrow \IC \IP^1 \times \IC$), this formalism yields the partition function as a sum over a single partition with Plancherel measure. In \cite{eynLP}, such a sum was rewritten as a 1-matrix integral. More complicated examples, such as the topological string on geometries underlying Seiberg-Witten $SU(n)$ theory, can be written as sums over multiple partitions \cite{Nekrasov, IK3, MarshakovNekrasov}. 1-matrix integrals that reproduce the corresponding partition functions were formulated in \cite{KlemmSulkowski}. Multi-matrix integrals have arisen in rewriting the framed vertex as a chain of matrices integral \cite{EynnPP}. Its Hurwitz-numbers limit (infinite framing of the framed vertex geometry) was shown to be reproduced by a 1-matrix model with an external field in \cite{Borot:2009ix, MorozovShakirov}.

Here, generalizing the method of \cite{eynLP}, we are able to formulate a matrix model which reproduces the topological string partition function on a certain fiducial geometry, which we introduce in the next section. Flop transitions and limits in the K\"ahler cone relate the fiducial geometry to an arbitrary toric Calabi-Yau manifold. As we can follow the effect of both of these operations on the topological string partition function, our matrix model provides a description for the topological string on an arbitrary toric Calabi-Yau manifold.

By providing a matrix model realization, we are able to transcribe deep structural insights into matrix models to the topological string setting. E.g., our matrix model involves a chain of matrices, and chain of matrices integrals are always tau functions for an integrable system. Our matrix model realization hence proves integrability of the generating function of Gromov-Witten invariants. Moreover, matrix models satisfy loop equations, which are known to be equivalent to W-algebra constraints. A general formal solution to these equations was found in \cite{Eynard:2005wg}, centered around the introduction of an auxiliary Riemann surface, referred to as the spectral curve of the system. The partition and correlation functions of the matrix model are identified with so-called symplectic invariants of this curve \cite{EOFg}. The BKMP conjecture \cite{BKMP}, building on work of \cite{marino2}, identifies the spectral invariants of the mirror curve to a toric Calabi-Yau manifold with the topological string partition function with the Calabi-Yau manifold as target space. In a forthcoming publication \cite{work_in_progress}, we will compute the spectral curve of our matrix model explicitly, thus establishing the validity of this conjecture.

Finally, we would like to emphasize that many different matrix models can yield the same partition function (justifying the choice of indefinite article in the title of this paper). An interesting open problem consists in identifying invariants of such equivalent matrix models. A promising candidate for such an invariant is the symplectic class of the matrix model spectral curve.

The outline of this paper is as follows. In section \ref{the_geometry}, after a very brief review of toric geometry basics, we introduce the fiducial geometry and the notation that we will use in discussing it throughout the paper. We also review the transformation properties of the topological string partition function under flop transitions, which will relate the fiducial to an arbitrary toric geometry, in this section. We recall the topological vertex formalism and its application to geometries on a strip \cite{IqbalKashaniPoor} in section \ref{vertex_calc}. Section \ref{our_matrix_model} contains our main result: we introduce a chain of matrices matrix model and demonstrate that it reproduces the topological string partition function on the fiducial geometry. By the argument above, we thus obtain a matrix model description for the topological string on an arbitrary toric Calabi-Yau manifold, in the large radius limit. We discuss implications of this result in section \ref{implications}, and point towards avenues for future work in section \ref{conclusions}.

\section{The fiducial geometry and flop transitions} \label{the_geometry}

Toric geometries present a rich class of very computable examples for many questions in algebraic geometry. The topological vertex formalism provides an algorithm for computing the generating function for Gromov-Witten invariants on toric 3 dimensional Calabi-Yau manifolds. These are necessarily non-compact and have rigid complex structure.

The geometry of toric manifolds of complex dimension $d$ can be encoded in terms of a $d$ dimensional fan $\Sigma$, consisting of cones of dimensions 0 to $d$. We denote the set of all $n$ dimensional cones as $\Sigma(n)$. Each such $n$-cone represents the closure of a $(C^*)^{d-n}$ orbit. In particular, 1-cones correspond to hypersurfaces, and for $d=3$, our case of interest, 2-cones correspond to curves.  

The fan for the class of geometries we are interested in is constructed by triangulating a finite connected region of the $\IZ^2$ lattice containing the origin, embedding this lattice in $\IZ^3$ within the $(x,y)$ plane at $z=1$, and defining the cones of the fan via half-lines emanating at the origin and passing through the vertices of this triangulation.\footnote{The canonical class of a toric manifold is given by the sum over all torically invariant divisors. The construction sketched above guarantees that this sum is principal, hence the canonical class trivial: the monomial associated to the 1-cone $(0,0,1)$ generates the class in question. See e.g.  \cite{Fulton}.}

\begin{figure}[h]
 \centering
  \includegraphics[width=5cm]{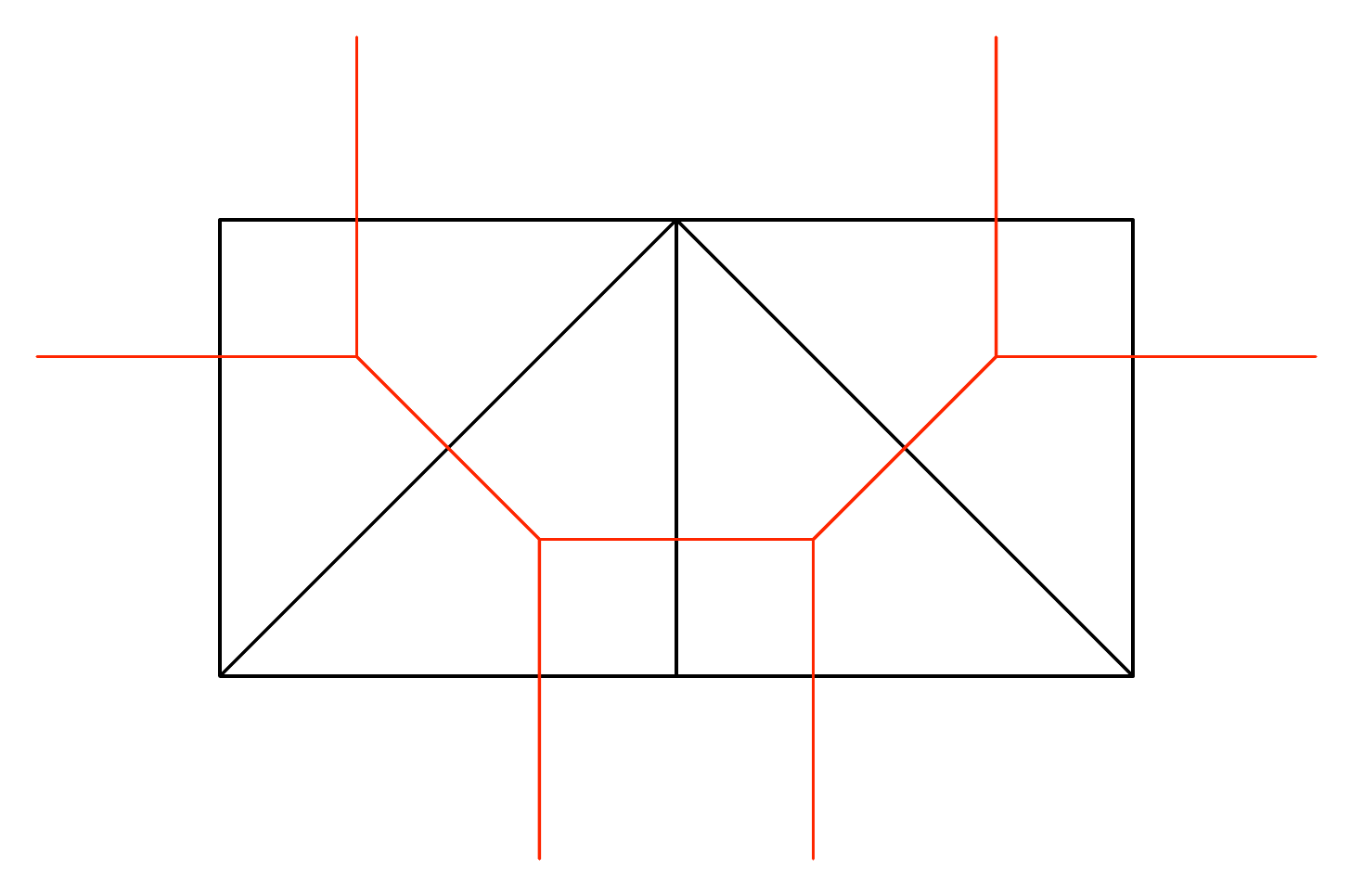}
 \caption{\footnotesize{Example of a box triangulation, corresponding to a 3 dimensional toric fan. The diagram in red is the dual web diagram. Vertices of the triangulations (faces of the web diagram) correspond to 1-cones, edges correspond to 2-cones, and faces (vertices of the dual) correspond to 3-cones.}}
 \label{web}
\end{figure}

We can associate a dual diagram to such toric fans, a so-called web diagram, spanned by lines orthogonal to the projection of 2-cones onto the $\IZ^2$ lattice. In the web diagram, the relation between the dimension of the components of the diagram and the submanifold of the toric geometry they represent coincide: 3-cones (points) correspond to vertices, and 2-cones (curves) to lines, see figure \ref{web}.

\subsection{The fiducial geometry}    \label{fiducial}
\begin{figure}[h]
\centering
\includegraphics[width=6cm]{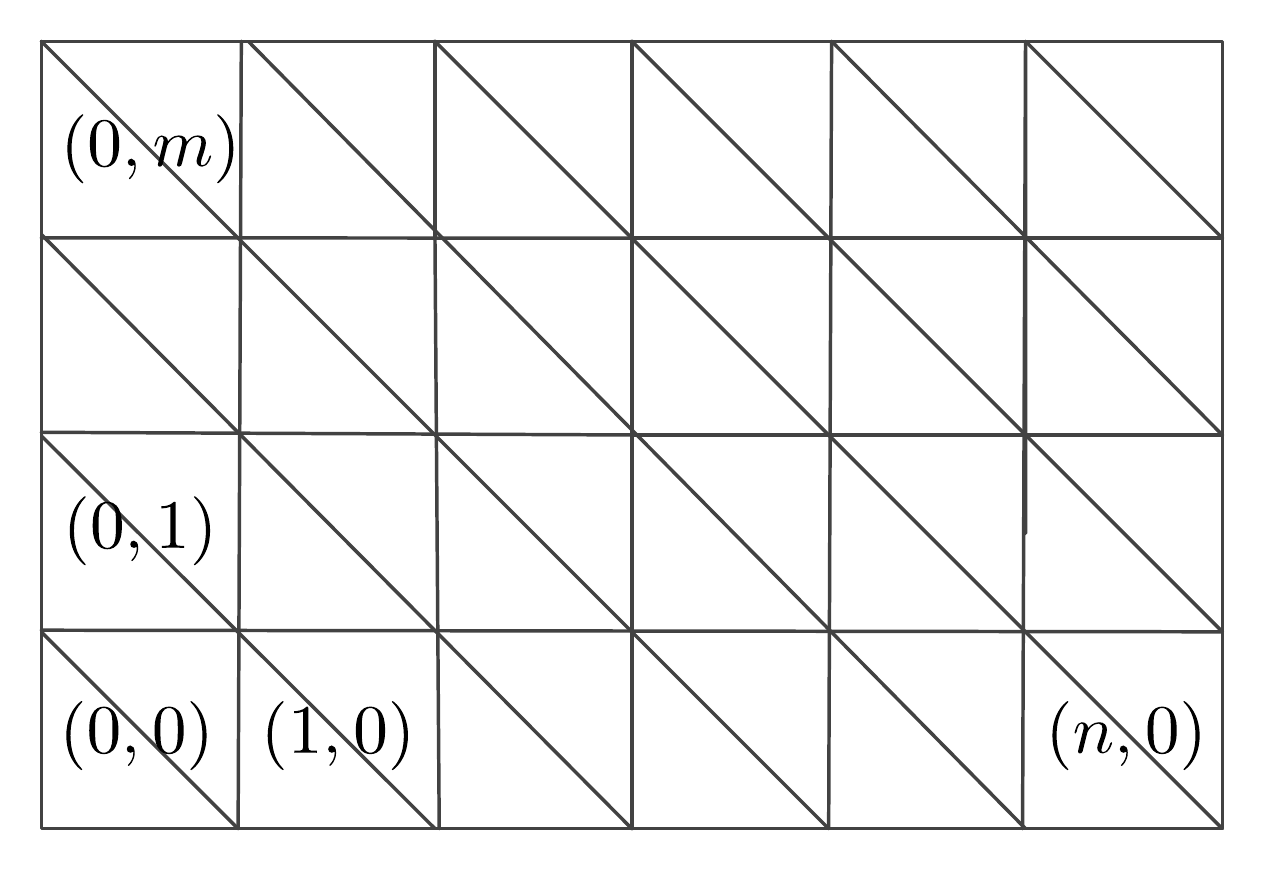}
\caption{\footnotesize{Fiducial geometry $\CYX_0$ with boxes numbered.}}
\label{fiducial_geometry_box}
\end{figure}

The geometry $\CYX_0$ we will take as the starting point of our considerations is depicted in figure \ref{fiducial_geometry_box}. 

\begin{figure}[h]
\centering
\includegraphics[width=3cm]{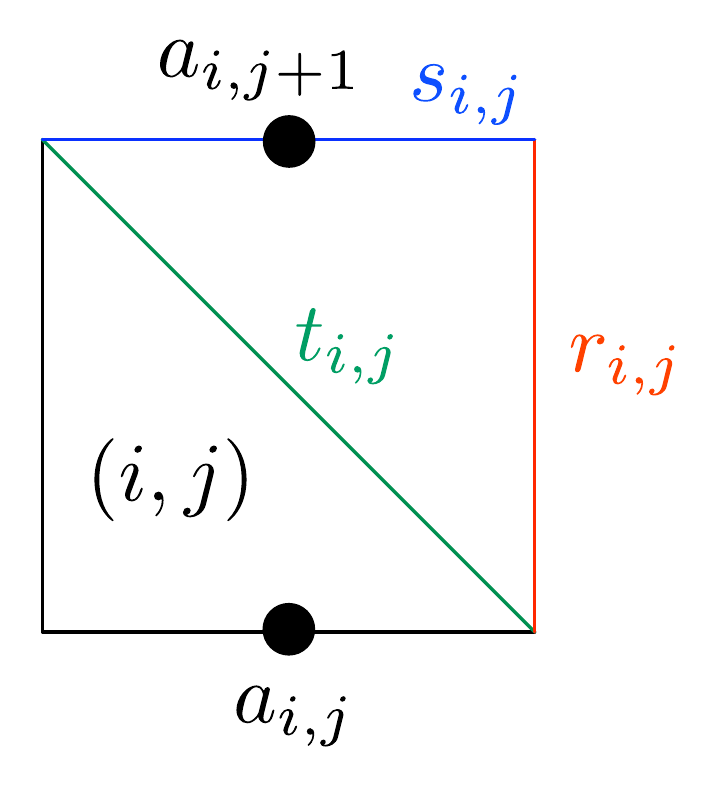}
\caption{\footnotesize{Labeling curve classes, and introducing $a$-parameters.}}
\label{fiducial_labeling}
\end{figure}

Since the torically invariant curves play a central role in our considerations, we introduce a labeling scheme for these in figure \ref{fiducial_labeling}: $(i,j)$ enumerates the boxes as in figure \ref{fiducial_geometry_box}, and we will explain the $a$-parameters further below.

\begin{figure}[h]
 \centering
 \includegraphics[width=8cm]{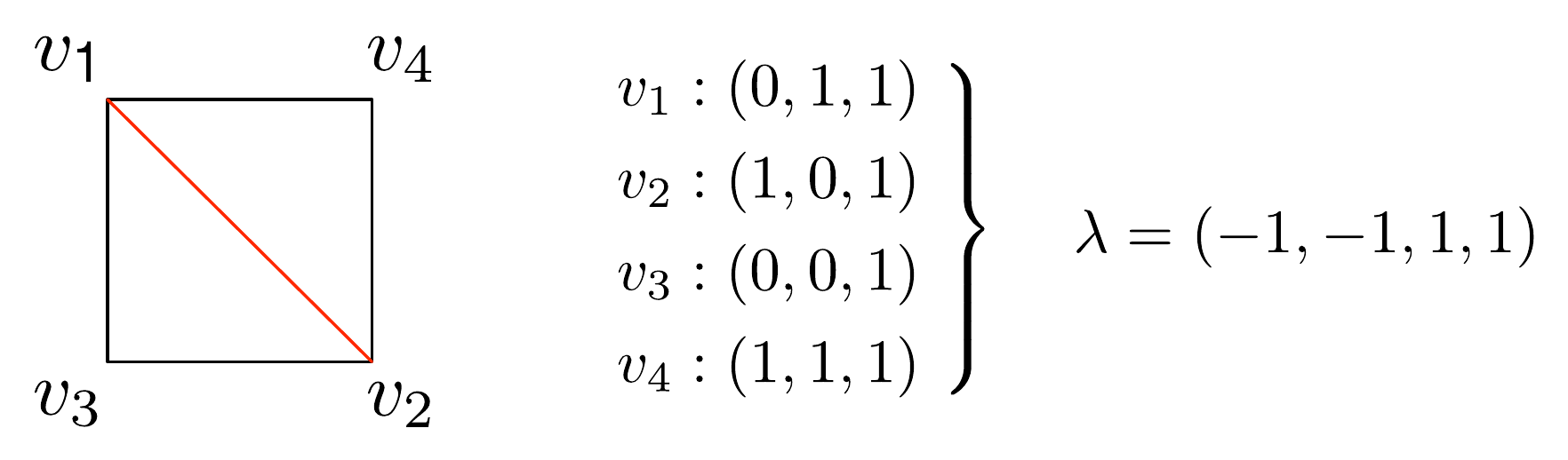}
 \caption{\footnotesize{The 2-cone $\sigma$ corresponds to the relation $\vec{\lambda}$ among 1-cones.}}
 \label{2conerelationcorr}
\end{figure}

In the following, we will, when convenient, use the same notation for a torically invariant curve $\Sigma$, its homology class $[\Sigma] \in H_2(\CYX_0,\IZ)$, and its volume or associated K\"ahler parameter $\int_\Sigma J$, given a K\"ahler form $J$ on $\CYX_0$. The classes of the curves $r_{i,j}, s_{i,j}, t_{i,j}$ introduced in figure \ref{fiducial_labeling} are not independent. To determine the relations among these, we follow \cite[page 39, 40]{CoxKatz}. Consider the integer lattice $\Lambda$ spanned by formal generators $e_\rho$, with $\rho \in \Sigma(1)$ 1-cones of the toric fan,
\beq
\Lambda = \{ \sum_{\rho \in \Sigma(1)} \lambda_\rho e_\rho | \lambda_\rho \in \IZ \} \,.
\eeq
Each torically invariant curve, corresponding to a 2-cone of the fan, maps to a relation between 1-cones, and thus to an element of the lattice $\Lambda$, as follows: a 2-cone $\sigma$ is spanned by two integral generators $v_1$ and $v_2$, and it is contained in precisely two 3-cones, which are each spanned by $v_1, v_2$ and one additional generator $v_3$, $v_4$ respectively. These vectors satisfy the relation $\sum_{i=1}^4 \lambda_{i} v_i =0$, where the $\lambda_{i}$ can be chosen as relatively prime integers, and as $v_3$ and $v_4$ lie on opposite sides of $\sigma$, we can assume that $\lambda_{3}, \lambda_{4} >0$. \cite{CoxKatz} shows that on a smooth variety, the sublattice $\Lambda_h$ generated by the elements $\sum_{i=1}^4 \lambda_{i} e_i$ of $\Lambda$ is isomorphic to $H_2(\CYX_0,\IZ)$. We call this isomorphism $\lambda$,
\beq 
\lambda : H_2(\CYX_0, \IZ) \rightarrow \Lambda_h \,.
\eeq
Figure \ref{2conerelationcorr} exemplifies this map.

\begin{figure}[h]
 \centering
  \includegraphics[width=4cm]{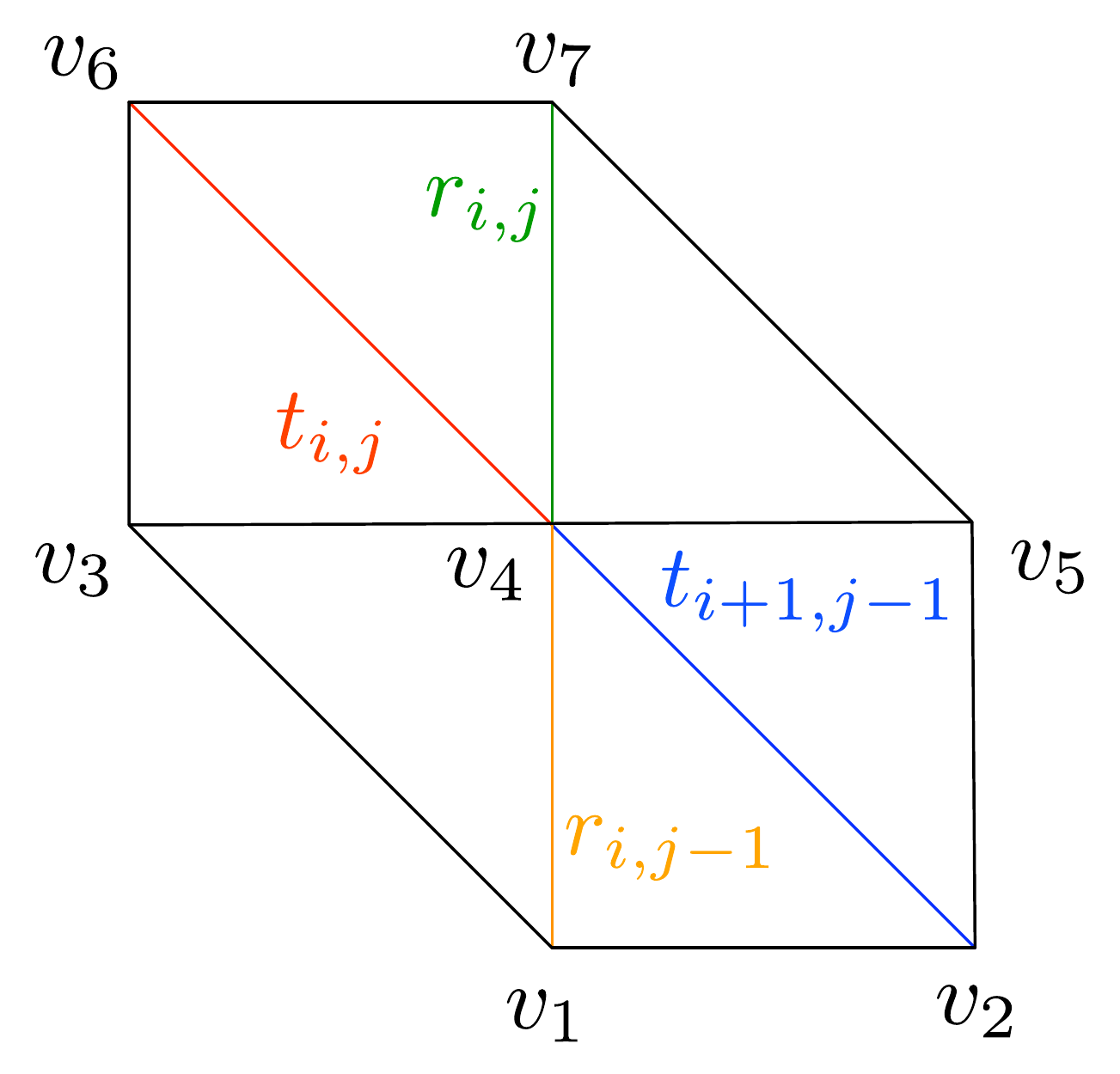}
 \caption{\footnotesize{Determining the relation between curve classes.}}
 \label{rel_bw_curve_classes}
\end{figure}

It allows us to easily work out the relation between the various curve classes. Consider figure \ref{rel_bw_curve_classes}.

\begin{figure}[h]
\centering
\includegraphics[width=12cm]{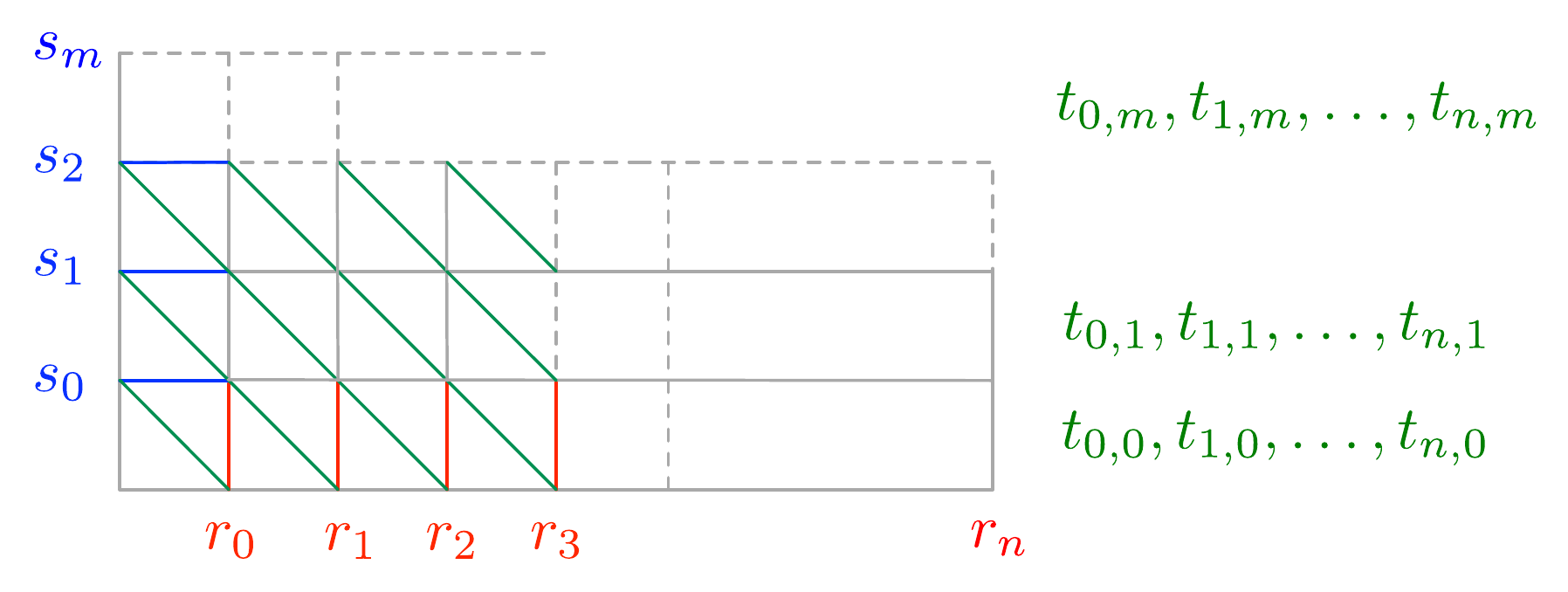}
\caption{\footnotesize{Fiducial geometry with choice of basis of $H_2(\CYX_0,\IZ)$.}}
\label{fiducial_geometry}
\end{figure}

The images of the curve classes depicted there under $\lambda$ are,
\ba
\lambda(r_{i,j}) &=& e_5 + e_6 - e_4 -e_7 \,,\\
\lambda(r_{i,j-1}) &=& e_2 + e_3 -e_1 -e_4\,,\\
\lambda(t_{i,j}) &=& e_3 + e_7 - e_4 - e_6 \,,\\
\lambda(t_{i+1,j-1}) &=& e_1 + e_5 - e_2 - e_4\,.\\
\ea
We read off the relation
\beqn   \label{tr_relation}
t_{i,j}+r_{i,j} = t_{i+1,j-1} + r_{i,j-1} \,.
\eeqn
By symmetry, we also have
\beq
t_{i,j} + s_{i,j-1} = t_{i+1,j-1} + s_{i+1,j-1} \,.
\eeq
A moment's thought convinces us that this constitutes a complete basis for the space of relations. We can solve these in terms of the classes of the curves $r_i, s_i, t_{i,j}$, $i,j = 0,1,\ldots$ depicted in figure \ref{fiducial_geometry}, which hence generate $H_2(\CYX_0,\IZ)$.
The explicit relations are
\ba
r_{i,j} &=& r_i + \sum_{k=1}^j (t_{i+1,k-1} - t_{i,k})\,, \\
s_{i,j} &=& s_j + \sum_{k=1}^i ( t_{k-1,j+1} - t_{k,j})\,.\\
\ea

Our computation for the partition function on $\CYX_0$ will proceed by first considering the horizontal strips in the toric fan describing the geometry, as depicted in figure \ref{fiducial_geometry_box}, individually, and then applying a gluing algorithm to obtain the final result.

For each strip, we find it convenient to write the curve class $w_{IJ} \in H_2(\CYX_0,\IZ)$ of the curve extending between two 3-cones which we label by $I$ and $J$ (recall that 3-cones correspond to vertices in the dual web diagram), with $J$ to the right of $I$, as the difference between two parameters $a_I$ and $a_J$ associated to each 3-cone, 
\beqn   \label{a_parameters}
w_{IJ} = a_I - a_J \,.
\eeqn
We call these parameters, somewhat prosaically, $a$-parameters. It is possible to label the curve classes in this way due to their additivity along a strip. In terms of the notation introduced in figure \ref{fiducial_labeling}, we obtain\beq
t_{i,j}=a_{i,j}-a_{i,j+1} \quad\,, \quad \quad r_{i,j} = a_{i,j+1} - a_{i+1,j} \,.
\eeq

By invoking the relation (\ref{tr_relation}), we easily verify that upon gluing two strips, the curve class of a curve extending between two 3-cones $I$ and $J$ on the lower strip is equal to the class of the curve between the 3-cones $I'$ and $J'$ on the upper strip, where the cones $I$ and $I'$ are glued together, as are the cones $J$ and $J'$,
\beqn
w_{IJ} = w_{I'J'} \,. \label{identify_class}
\eeqn
This allows us to identify the parameters $a_I=a_{I'}$ and $a_J=a_{J'}$ associated to 3-cones glued together across strips.

Note that the basic curve classes $s_{i}$ are not captured by the parameters $a_{i,j}$.

\subsection{Flop invariance of toric Gromov-Witten invariants} \label{flops}

\begin{figure}[h]
 \centering
 \includegraphics[width=6cm]{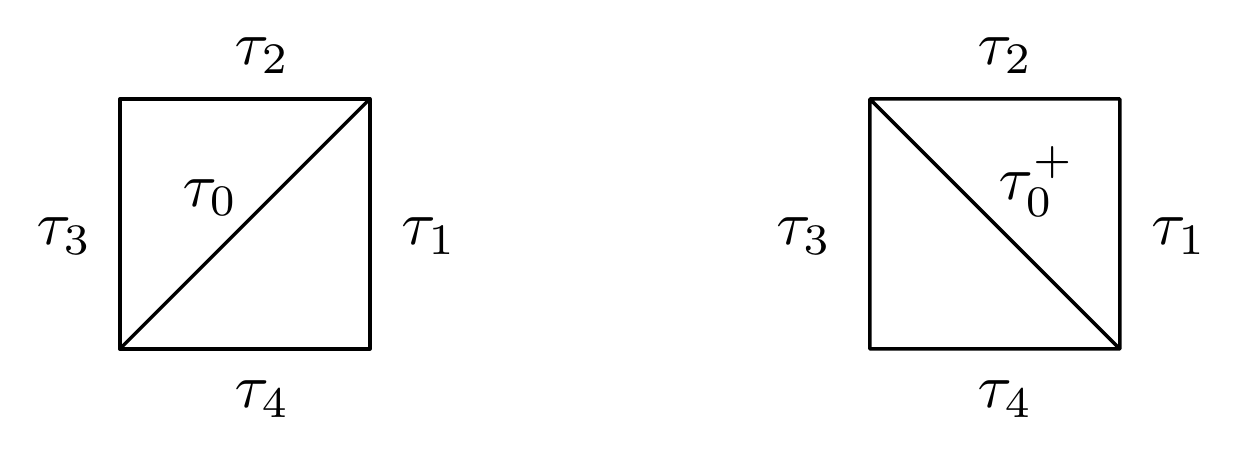}
 \caption{\footnotesize{$\CYX$ and $\CYX^+$ in the vicinity of the (-1,-1) curve.}}
 \label{toric_flopped}
\end{figure}

Under the proper identification of curve classes, Gromov-Witten invariants (at least on toric manifolds) are invariant under flops. Assume $\CYX$ and $\CYX^+$ are related via a flop transition, $\phi: \CYX \rightarrow \CYX^+$. In a neighborhood of the flopped $(-1,-1)$ curve, the respective toric diagrams are depicted in figure \ref{toric_flopped}.

The 1-cones of $\Sigma_\CYX$, corresponding to the toric invariant divisors of $\CYX$, are not affected by the flop, hence can be canonically identified with those of $\CYX^+$. The 2-cones $\tau_i$ in these diagrams correspond to toric invariant 2-cycles $C_i$, $C_i^+$ in the geometry. The curve classes of $\CYX$ push forward to classes in $\CYX^+$ via
\ban\label{flopping}
\phi_*([C_0]) = - [C_0^+] \,, \quad \phi_*([C_i]) =  [C_i^+] + [C_0^+] \,. 
\ean
All other curve classes of $\CYX$ are mapped to their canonical counterparts in $\CYX^+$. Under appropriate analytic continuation and up to a phase factor (hence the $\propto$ in the following formula), the following identity then holds \cite{Witten_Phases,IqbalKashaniPoor, KonishiMinabe},
\ban
Z_{GW}(\CYX,Q_0,Q_1,\ldots,Q_4,\vec{Q}) \propto Z_{GW}(\CYX^+,1/Q_0, Q_0 Q_1, \ldots, Q_0 Q_4,\vec{Q}) \,,  \label{rel_part_flopped}
\ean
i.e.
\ba
GW_g(\CYX,Q_0,Q_1,\ldots,Q_4,\vec{Q}) = GW_g(\CYX^+,1/Q_0, Q_0 Q_1, \ldots, Q_0 Q_4,\vec{Q}) \,.
\ea

Any toric Calabi-Yau manifold $\CYX$ with K\"ahler moduli $\vec{Q}$ can be obtained from a sufficiently large fiducial geometry $(\CYX_0,\vec{Q}_0)$ upon performing a series of flop transitions and taking unwanted K\"ahler moduli of $\CYX_0$ to $\infty$. Once we obtain a matrix model reproducing the topological string partition function on the fiducial geometry, extending the result to arbitrary toric Calabi-Yau 3-folds will therefore be immediate.

\begin{figure}[h]
 \centering
 \includegraphics[width=10cm]{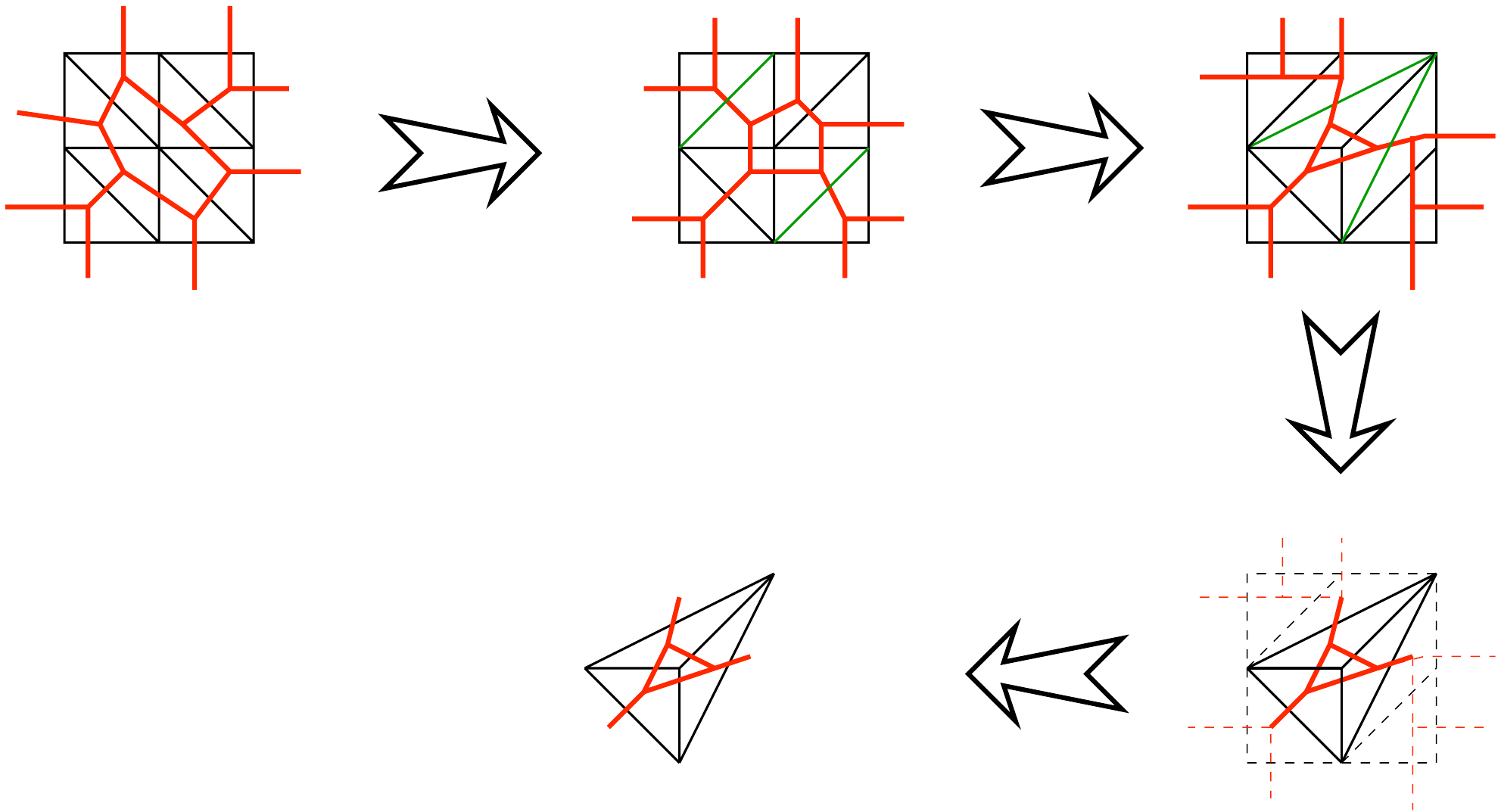}
 \caption{\footnotesize{We obtain local $\mathbb P^2$ from the fiducial geometry with $2\times 2$ boxes by performing five flops and then sending the K\"ahler parameters of the unwanted edges to $\infty$.}}
 \label{flopP2}
\end{figure}

As an example, we show how to obtain the $\mathbb P^2$ geometry from the fiducial geometry with $2\times 2$ boxes in figure \ref{flopP2}.

\section{The partition function via the topological vertex} \label{vertex_calc}
\subsection{Gromov-Witten invariants}

Gromov-Witten invariants ${\cal N}_{g,D}(\CYX)$ roughly speaking count the number of maps from a Riemann surface of genus $g$ into the target space $\CYX$, with image in a given homology class $D=(D_1,\dots,D_{k})\in H_2(\CYX,\mathbb Z)$. They can be assembled into a generating series
\beq
GW_g(\CYX,Q) = \sum_{D} {\cal N}_{g,D}(\CYX)\, Q^{D}.
\eeq
Each $GW_g(\CYX,Q)$ is a formal series in powers $Q^D = \prod_i Q_i^{D_i}$ of the parameters $Q=(Q_1,Q_2,\dots,Q_k)$, the exponentials of the K\"ahler parameters. 

\medskip

We can introduce a generating function for Gromov-Witten invariants of all genera by introducing a formal parameter $g_s$ (the string coupling constant) and writing
\beq
GW(\CYX,Q,g_s) = \sum_{g=0}^\infty\,\, g_s^{2g-2}\,\, GW_g(\CYX,Q) \,.
\eeq
%This equality is an equality between formal power series of $Q$, and to each order, the sum over $g$ is finite.

\bigskip

It is in fact more convenient to introduce disconnected Gromov-Witten invariants  ${\cal N}^*_{\chi,D}(\CYX)$, for possibly disconnected surfaces, of total Euler characteristics $\chi$, and to define
\beq
Z_{GW}(\CYX,Q,g_s) = e^{GW(\CYX,Q,g_s)} =  \sum_{D} \, Q^D\,\, \sum_\chi  g_s^{-\chi} \,\, {\cal N}^*_{\chi,D}(\CYX).
\eeq

For toric Calabi-Yau manifolds, an explicit algorithm was presented in \cite{AKMV} for computing $Z_{GW}$ via the so-called topological vertex formalism, proved in \cite{LLLZ,MOOP}.

\subsection{The topological vertex}

In the topological vertex formalism, each vertex of the web diagram contributes a factor $C_q({\alpha, \beta, \gamma})$ to the generating function of GW-invariants, 
where the $\alpha, \beta, \gamma$ are Young tableaux associated to each leg of the vertex, 
and $C_q({\alpha, \beta, \gamma})$  is a formal power series in the variable $q$, where
$$
q=e^{-g_s}.
$$
Topological vertices are glued along edges (with possible framing factors, see \cite{AKMV}) carrying the same Young tableaux $\alpha$ by performing a sum over $\alpha$, weighted by $Q^{|\alpha|}$, with $Q$ encoding the curve class of this connecting line,
\beq
Z_{\rm vertex}(\CYX,Q,q)=\sum_{{\rm Young\,\, tableaux}\,\alpha_e}\,\, \prod_{{\rm edges}\, e}\, Q_e^{|\alpha_e|}\,\,\, \prod_{{\rm vertices}\, v =(e_1,e_2,e_3)}\,\, C_q(\alpha_{e_1},\alpha_{e_2},\alpha_{e_3}) \,.
\eeq
Note that in practical computations, the sum over representations can ordinarily not be performed analytically. A cutoff on the sum corresponds to a cutoff on the degree of the maps being counted.
 
The equality
\beq
Z_{GW}(\CYX,Q,g_s) = Z_{\rm vertex}(\CYX,Q,q)
\eeq
holds at the level of formal power series in the $Q$'s, referred to as the large radius expansion. It was proved in \cite{MOOP} that the log of the right hand side indeed has a power series expansion in powers of $g_s$.

\subsection{Notations for partitions and q-numbers}

Before going further in the description of the topological vertex formula, we pause to fix some notations and introduce special functions that we will need in the following.

\subsubsection{Representations and partitions} \label{secrepresentations}

Representations of the symmetric group are labelled by Young tableaux, or Ferrer diagrams. For a representation $\gamma$, we introduce the following notation:
\begin{itemize}
\item $\gamma_i$: number of boxes in the $i$-th row of the Young tableau associated to the representation $\gamma$, $\gamma_1\geq \gamma_2\geq \dots \geq \gamma_{d}\geq 0$.

\item The weight $|\gamma|=\sum_i \gamma_i$: the total number of boxes in the corresponding Young tableau.
\item The length $l(\gamma)$: the number of non-vanishing rows in the Young tableau, i.e. $\gamma_i=0$ iff $i>l(\gamma)$.

\item The Casimir $\kappa(\gamma) = \sum_i \gamma_i (\gamma_i - 2i +1 )$.

\item $\gamma^T$ denotes the conjugate representation, which is obtained by exchanging the rows and columns of the associated Young tableau. We have $|\gamma^T|=|\gamma|$, $l(\gamma^T)=\gamma_1$, and $\kappa(\gamma^T)=-\kappa(\gamma)$.

\end{itemize}

\medskip

An integer $d>0$ will denote a cut-off on the length of representations summed over, 
\ba
l(\gamma)\leq d.
\ea
Most expressions we are going to write will in fact be independent of $d$, and we shall argue in \cite{work_in_progress}, following the same logic as in \cite{eynLP} based on the arctic circle property \cite{Johansson}, that our results depend on $d$ only non-perturbatively.

\smallskip
To each representation $\gamma$, we shall associate a parameter $a$ as introduced in (\ref{a_parameters}).

\begin{figure}[h]
 \centering
  \includegraphics[width=14cm]{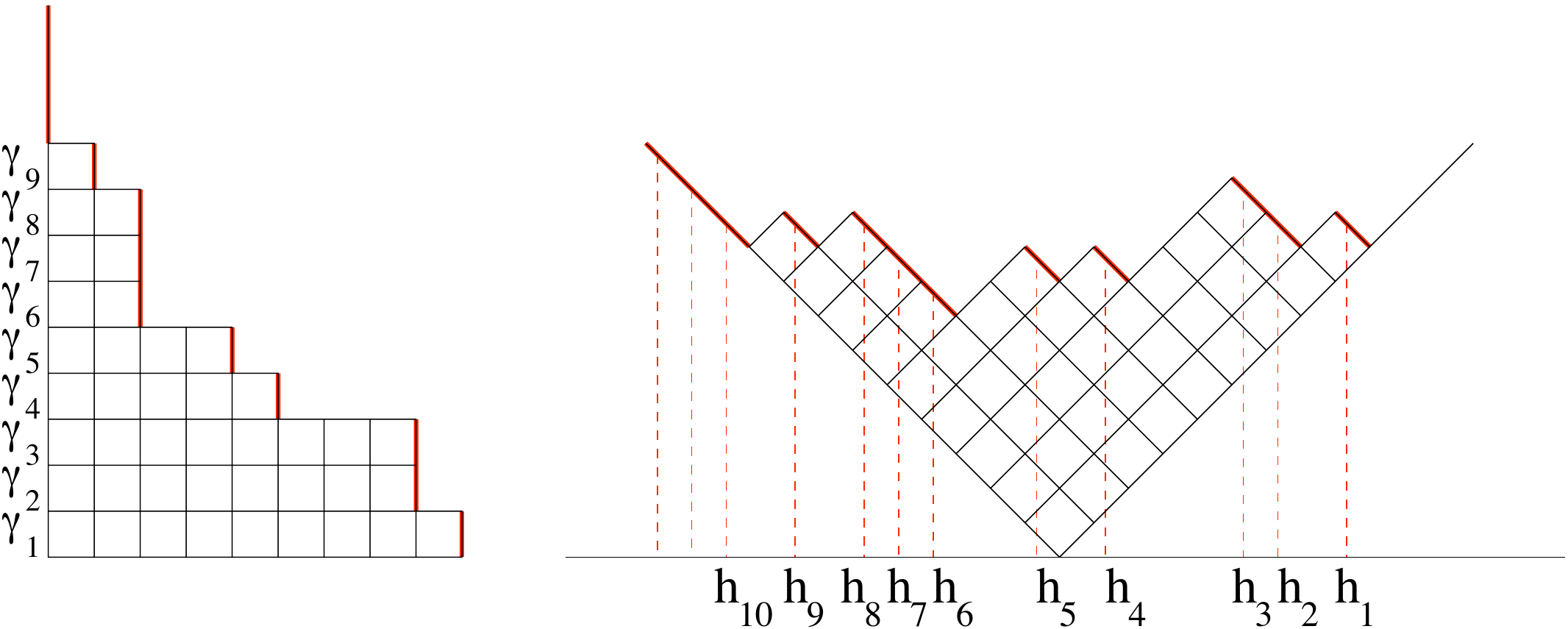}
 \caption{\footnotesize{Relation between a partition $\gamma$ and $h(\gamma)$.}}
 \label{expartition}
\end{figure}

Instead of dealing with a partition $\gamma$, characterized by the condition $\gamma_1 \ge \gamma_2 \ge \ldots \ge \gamma_d \ge 0$, it will prove convenient to define the quantities
\beqn \label{defh}
h_i(\gamma)=\gamma_i-i+d+a \,,
\eeqn
which satisfy instead
\beq
h_1>h_2>h_3>\dots >h_{d}\geq a  \,.
\eeq
The relation between $\gamma$ and $h(\gamma)$, for the off-set $a=0$, is depicted in figure \ref{expartition}.

We finally introduce the functions
\beq 
x_i(\gamma)=q^{h_i(\gamma)}  \,.
\eeq

In terms of the $h_i(\gamma)$, we have
\ba
\kappa(\gamma)=\sum_i h_i^2 - (2d+2a-1)\sum_i h_i + d\,C_{d,a} \,,
\ea
where $C_{d,a}=\frac{1}{3}(d-1)(2d-1)+\,a(a+2d-1)$.

\subsubsection{q-numbers}

We choose a string coupling constant $g_s$ such that the quantum parameter
$q= e^{-g_s}$ satisfies $|q|<1$. A $q$-number $[x]$ is defined as
\beqn  \label{q_number}
[x] = q^{-\frac{x}{2}} - q^{\frac{x}{2}} = 2 \sinh \frac{x\,g_s}{2} \,.
\eeqn
$q$-numbers are a natural deformation away from the integers; in the limit $q\to 1$, ${1\over g_s}\,[x] \rightarrow  x$.

We also define the $q$-product
\ba
g(x) = \prod_{n=1}^\infty (1-{1\over x}\,q^n) \,.
\ea
The function $g(x)$ is related to the quantum Pochhammer symbol, $g(x)
= [q/x;q]_{\infty}$, and to the $q$-deformed gamma function via $\Gamma_q(x) = (1-q)^{1-x}\,g(1)/g(q^{1-x})$.
$g(x)$ satisfies the functional relation
\ba
g(qx) = (1-{1\over x})\,\, g(x) \,.
\ea
For $\Gamma_q$, this implies $\Gamma_q(x+1) = {1-q^x\over 1-q}\,\, \Gamma_q(x)$, the
quantum deformation of the functional equation $\Gamma(x+1)=x\Gamma(x)$ of the gamma function, which is recovered in the classical limit $q\to 1$.
The central property of $g(x)$ for our purposes is that it vanishes on integer powers of $q$,
\beq
g(q^n)=0 \quad {\rm if}\,\, n\in \mathbb N^*.
\eeq
Moreover, it has the following small $\ln q$ behavior,
\beq
\ln{g(x)} = {1\over \ln q}\,\sum_{n=0}^\infty {(-1)^n\,B_n\over n!}\,(\ln q)^n\,\, \Li_{2-n}(1/x)  \,,
\eeq
where $\Li_n(x) = \sum_{k=1}^\infty {x^k\over k^n}$ is the polylogarithm, and $B_n$ are the Bernouilli numbers
\ba
B_0=1 \,\, , \quad B_1 =-{1\over 2}\,\, , \quad B_2={1\over 6}\,\, , \quad \dots
\ea
$B_{2k+1}=0$ if $k\geq 1$ (see the appendix).

We shall also need the following function $f(x)$,
\ba
{1\over f(x)} 
&=& {g(x) \, g(q/x)\over g(1)^2\,\,\sqrt x}\,\,e^{(\ln{x})^2\over 2\ln q}\,\, e^{-i\pi \ln x\over \ln q}   
%= {e^{-i\pi\ln x\over \ln q}\over \Gamma_q(q/x)\,\,\Gamma_q(x)} 
\cr
&=& {-\ln q\over \theta'({1\over 2} - {i\pi\over \ln q}, -{2i\pi\over \ln q}) }\,\,\, \theta\left({\ln x\over \ln q} +{1\over 2} - {i\pi \over \ln q}, {-2i\pi\over \ln q}\right) ,
\ea
where $\theta$ is the Riemann theta-function for the torus of modulus $-2i\pi/\ln q$.
This relationship is the quantum deformation of the classical gamma function identity $$e^{-i\pi x}/\Gamma(1-x)\Gamma(x) = \sin{(\pi x)}/\pi \,.$$

\subsection{The partition function via the vertex\label{secZvertex}}

\begin{figure}[h]
 \centering 
  \includegraphics[width=15cm]{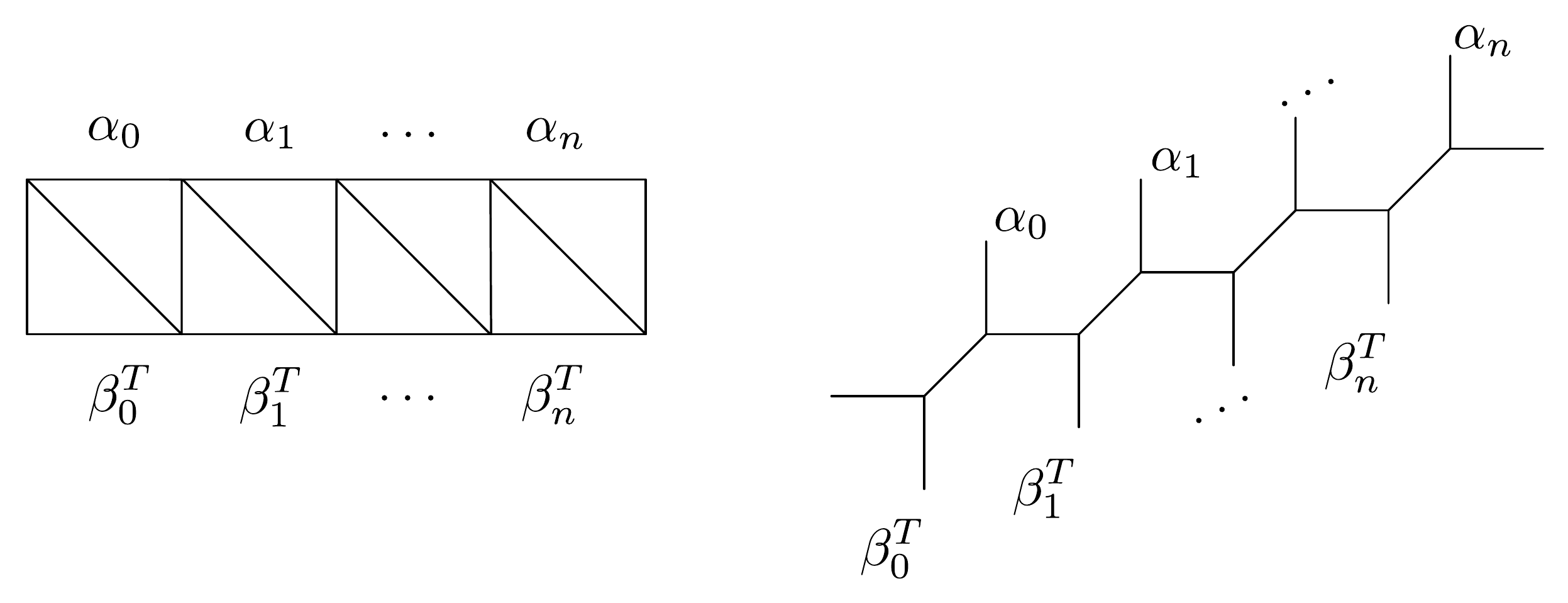}
 \caption{\footnotesize{A horizontal strip of the fiducial geometry and its corresponding web diagram. \label{figstrip}}}
\end{figure}

We begin by considering a single horizontal strip of the fiducial geometry, as depicted in figure \ref{figstrip}.

Of the three legs of the vertex, two point in the direction of the strip and connect the vertex to its neighbors. One leg points out of the strip, either above or below. This leg carries a free representation, $\alpha_i$ or $\beta_i^T$ in the notation of figure \ref{figstrip}. The partition function will hence depend on representations, one per vertex (i.e. face of the triangulation). 

A note on notation: since each 3-cone carries a representation (which up to the final paragraph of this subsection is held fixed) and an a-parameter (see figure \ref{fiducial_labeling}), we will identify the a-parameters by the corresponding representations when convenient.

Using the topological vertex, it was shown in \cite{IqbalKashaniPoor} that the A-model topological string partition function of the strip is given by a product of terms, with the individual factors depending on the external representations and all possible pairings of these. Applied to the fiducial strip, the results there specialize to
\ban
Z_{\rm strip}(\alpha_0;\beta^T)  \label{z_strip}
&=&   \prod_{i=0}^n \frac { [\alpha_i] [\beta^T_i] }{\,\quad \quad [\beta_i,\alpha_i^T]_{Q_{\beta_i,\alpha_i}}} \,\,\,\, {\prod_{i<j} [\alpha_i,\alpha_j^T]_{Q_{\alpha_i,\alpha_j}} \,\,\prod_{i<j} [\beta_i,\beta_j^T]_{Q_{\beta_i,\beta_j}} 
\over \prod_{i<j} [\alpha_i,\beta_j^T]_{Q_{\alpha_i,\beta_j}} [\beta_i,\alpha_j^T]_{Q_{\beta_i,\alpha_j}}
}    
\label{pfcn} \,.
\ean
We explain each factor in turn.

\medskip

$\bullet$ Each vertex $\gamma=\alpha_i$ or $\gamma=\beta_i^T$ contributes a representation dependent factor to the partition function, which we have denoted by $[\gamma]$. It is the $n \rightarrow \infty$ limit of the Schur polynomial evaluated for $x_i = q^{\frac{1}{2}-i}$, $i=1, \ldots, n$, given explicitly by
\ba
[\gamma] 
&=& (-1)^d  q^{\frac{1}{4} \kappa(\gamma)} \prod_{1 \le i < j \le d} \frac{[\gamma_i - \gamma_j + j - i]}{[j-i]} \prod_{i=1}^{d} \prod_{j=1}^{\gamma_i} \frac{1}{[d + j - i]} \\
&=&
\prod_{1\leq i< j\leq d} (q^{h_j}-q^{h_i}) \,\,\prod_{i=1}^d \, \left(\frac{ g(q^{a_\gamma-h_i})}{g(1)}\,\,q^{\frac{1}{2}h_i^2-(a_\gamma+d-1)h_i+\frac{a_\gamma(a_\gamma+d-1)}{2}+\frac{(d-1)(2d-1)}{12}}\right)  \cr
&=& \Delta(X(\gamma))\,\,  e^{-{1\over g_s} \tr U(X(\gamma),a_\gamma)}\,\,e^{-{1\over g_s} \tr U_1(X(\gamma),a_\gamma)}.
\ea
We recall that $h_i(\gamma) = \gamma_i-i+d+a_\gamma$, and we have defined $x_i=q^{h_i}$ and the diagonal matrix $X(\gamma)={\rm diag}(q^{h_1},q^{h_2},\dots,q^{h_d})$. Furthermore, $\Delta(X)$ denotes the Vandermonde determinant of the matrix $X$,
\beq
\Delta(X) = \prod_{1\leq i<j\leq d} (x_j-x_i) \,,
\eeq
and we have written
\beq
U(X,a)  = - g_s \ln{\left( g({q^a\over X})\over g(1)\right)}  \,,
\eeq
\beq
U_1(X,a) =  {(\ln X)^2\over 2} - (a+d-1)\ln{X}\,\ln q  + C(a,d) \,,
\eeq
where $C(a,d) = \frac{a(a+d-1)}{2}+\frac{(d-1)(2d-1)}{12}$.

We have
$$ [\gamma] = q^{\frac{\kappa(\gamma)}{2}} [\gamma^T]\,, \quad \kappa(\gamma^T) = - \kappa(\gamma) \,,$$
and thus
\ba
[\gamma^T] 
&=& \Delta(X(\gamma))\,\,  e^{-{1\over g_s} \tr U(X(\gamma),a_\gamma)}\,\,e^{-{1\over g_s} \tr \td U_1(X(\gamma),a_\gamma)} \,,
\ea
where
\beq
\td U_1(X,a)  = {1\over 2} \ln{X}\,\ln q  + \td C(a,d).
\eeq
$\td C_{a,d}$ is another constant which depends only on $a$ and $d$ and which will play no role for our purposes.

\medskip

$\bullet$ 
In addition, each pair of representations contributes a factor, reflecting the contribution of the curve extended between the respective vertices. In the nomenclature of \cite{IqbalKashaniPoor}, the representations $\alpha_i$ are all of same type, and of opposite type relative to the $\beta_i$. If we take $i<j$, representations of same type (corresponding to (-2,0) curves) contribute a factor of $$[\alpha_i,\alpha_j^T] \quad \mbox{or} \quad [\beta_i^T,\beta_j]\,,$$ whereas representations of different type (corresponding to (-1,-1) curves) contribute a factor of $$\frac{1}{[\alpha_i,\beta_j]}  \quad \mbox{or} \quad  \frac{1}{[\beta_i^T,\alpha_j^T]}\,.$$
The pairing is given by \cite{IK2,IK3,EguchiKanno,IqbalKashaniPoor}
\ban  \label{pairingproduct}
[\gamma, \delta^T]
&=&  Q_{\gamma,\delta}^{-\frac{|\gamma| + |\delta|}{2}} q^{-\frac{\kappa(\gamma) - \kappa(\delta) }{4}} \prod_{i=1}^{d} \prod_{j=1}^{d} \frac{  [ h_i(\gamma)-h_j(\delta)] }{ [ a_\gamma-a_\delta  + j - i] }  \nn\\
&& \times \prod_{i=1}^{d} \prod_{j=1}^{\gamma_i} \frac{1}{[a_\gamma-a_\delta + j - i + d]} \prod_{i=1}^{d} \prod_{j=1}^{\delta_i} \frac{1}{[a_\gamma-a_\delta - j + i - d]} \,\,  \prod_{k=0}^\infty g(Q_{\gamma,\delta}^{-1}q^{-k})\, \nn\\
&=& (-1)^{\frac{d(d-1)}{2}} \prod_{i=1}^d \frac{q^{\frac{1}{2} (h_i(\delta)^2 - h_i(\delta) (2 a_\gamma +2d -1) - a_\delta^2 + 2 a_\gamma a_\delta + (d-2i)a_\gamma + (2i - d - 1) a_\delta )}}{[a_\gamma - a_\delta]^{d}}  \prod_{i=1}^d (-1)^{\delta_i}  \nn \\
&&\prod_{i,j=1}^d (q^{h_j(\delta)}- q^{h_i(\gamma)}) \prod_{i=1}^d \frac{g(q^{a_\gamma-h_i(\delta)})}{g(q^{a_\gamma-a_\delta})} \frac{g(q^{a_\delta-h_i(\gamma)})}{g(q^{a_\delta-a_\gamma})} \nn \\
%&\propto& \Delta(X(\gamma),X(\delta))\,\, 
%\prod_i g(q^{a_\delta-h_i(\gamma)}) \,\, \prod_i g(q^{a_\gamma-h_i(\delta)}) \, q^{{1\over 2}\,h_i(\delta)^2}\,\, q^{-{h_i(\delta)\over 2}(2a_\gamma+2d-1)} \cr
&\propto& \Delta(X(\gamma),X(\delta))\,\, e^{-{1\over g_s} \Tr U(X(\gamma),a_\delta)} e^{-{1\over g_s} \Tr U(X(\delta),a_\gamma)} \,\, e^{-{1\over g_s}\left( \Tr U_2(X(\gamma),a_\delta)+ \Tr \td U_2(X(\delta),a_\gamma)\right)}\, , \nn \\
\ean
where the square brackets on the RHS denote $q$-numbers as defined in (\ref{q_number}), the symbol $\Delta(X(\gamma),X(\delta))$ signifies
\ban \label{doublevan}
\Delta(X(\gamma),X(\delta)) = \prod_{i,j} (X_i(\delta)-X_j(\gamma)) = \prod_{i,j} (q^{h_i(\delta)}-q^{h_j(\gamma)}) \,,
\ean
and
\ba
U_2(X,a)=0 \,,
\ea
\ba
\td U_2(X,a)= {(\ln X)^2\over 2} - (a+d-{1\over 2})\, \ln X\,\ln q + i\pi \ln X.
\ea
The parameter $Q_{\gamma, \delta}$ reflects, given a choice of K\"ahler class $J$ of the metric on $\CYX_0$, the curve class of the curve $\cC$ extended between the vertices labeled by $\gamma$ and $\delta$ via
\beq
w_{\gamma,\delta} = \int_{\cC} J \, \,,\quad \quad Q_{\gamma, \delta} = q^{w_{\gamma,\delta}} \,.
\eeq
By the definition of the a-parameters,
\beq
w_{\gamma,\delta} = a_\gamma-a_\delta.
\eeq

Substituting these expressions into (\ref{z_strip}), we obtain
\ban
\lefteqn{Z_{\rm strip}(\alpha_0,\dots,\alpha_n;\beta_0^T,\dots,\beta_n^T) =}\cr
&& \cr
&=& {\prod_i \Delta(X(\alpha_i))\,\prod_{i<j}\Delta(X(\alpha_i),X(\alpha_j))\,\,\, \prod_i \Delta(X(\beta_i))\,\prod_{i<j}\Delta(X(\beta_i),X(\beta_j))
\over \prod_{i,j} \Delta(X(\alpha_i),X(\beta_j))} \cr
&& \times \prod_i e^{-{1\over g_s}\, \tr (V_{\vec a}(X(\alpha_i))-V_{\vec b}(X(\alpha_i)))}\,\,  \prod_i e^{-{1\over g_s}\, \tr V_i(X(\alpha_i))} \cr
&&\times \prod_i e^{{1\over g_s}\, \tr  (V_{\vec a}(X(\beta_i))-V_{\vec b}(X(\beta_i)))}\,\,  \prod_i e^{-{1\over g_s}\, \tr \td V_i(X(\beta_i))}  \,,  \label{Zstripsum1}
\ean
where we have denoted by $\vec a = (a_0,a_1,\dots,a_n)$ (resp. $\vec b = (b_0,b_1,\dots,b_n)$) the a-parameters of representations on the upper side (resp. lower side) of the strip, and defined
\beqn    \label{partpot}
V_{\vec a}(X) = -g_s\,\sum_{j=0}^n \ln{\left(g(q^{a_j}/X)\right)} \,,
\eeqn
and
\beq
V_i(X) = \ln X\, \ln q\,\, \left({1\over 2} - \sum_{j\leq i} (a_j-b_j)\right)+i\pi \ln X \,,
\eeq
\beq
\td V_i(X) = \ln X\, \ln q\,\, \left({1\over 2} - \sum_{j<i} (b_j-a_j)\right) \,.
\eeq

\subsection{Gluing strips}

To obtain the partition function for the full multistrip fiducial geometry $\CYX_0$, we must glue these strips along the curves labelled $s_{i,j}$ in figure \ref{fiducial_labeling}.

Denoting the representations $\alpha_{j,i}$ on line $i$ collectively by
\beq
\vec\alpha_i = (\alpha_{0,i},\alpha_{1,i},\dots,\alpha_{n,i})  \,,
\eeq
this yields
\ban
Z_{\rm vertex}(\CYX_0) = Z_{(n,m)}(\vec\alpha_{m+1},\vec\alpha_0^T) 
=  \sum_{\alpha_{j,i},\, j=0,\dots,n;\, i=1,\dots,m} \quad \,\,
\prod_{i=1}^{m+1} Z_{\rm strip}(\vec\alpha_i,\vec\alpha_{i-1}^T) \,\, \prod_{j=0}^n\prod_{i=1}^m q^{s_{j,i}\,|\alpha_{j,i}|} \,. \nonumber\\  \label{ZvertexprodZstrips}
\ean

Our goal now is to find a matrix integral which evaluates to this sum.

\section{The matrix model}   \label{our_matrix_model}

\subsection{Definition}

Consider the fiducial geometry $\CYX_0$ of size $(n+1)\times (m+1)$, with K\"ahler parameters $t_{i,j}=a_{i,j}-a_{i,j+1}$, $r_{i,j} = a_{i,j+1}-a_{i+1,j}$, and  $s_{i,j}$, as depicted in figures \ref{fiducial_labeling} and \ref{fiducial_geometry}. We write
\beq
\vec a_i  = (a_{0,i},a_{1,i},\dots,a_{n,i}).
\eeq

Assume that the external representations are fixed to $\vec\alpha_{m+1} = (\alpha_{0,m+1},\alpha_{1,m+1},\dots,\alpha_{n,m+1})$ on the upper line, and $\vec\alpha_0 = (\alpha_{0,0},\alpha_{1,0},\dots,\alpha_{n,0})$ on the lower line 
(for most applications, one prefers to choose these to be trivial).

\smallskip

We now define the following matrix integral ${\cal Z}_{\rm MM}$ (${}_{\rm MM}$ for Matrix Model),
\ban
{\cal Z}_{\rm MM}(Q,g_s,\vec\alpha_{m+1},\vec\alpha_0^T)
&=& \Delta(X(\vec \alpha_{m+1}))\,\, \Delta(X(\vec \alpha_0)) \,\, 
\prod_{i=0}^{m+1} \int_{H_N(\Gamma_i)} dM_i \,
 \prod_{i=1}^{m+1}\int_{H_N({\mathbb R}_+)}\,dR_i \nn \\
&& \prod_{i=1}^{m} e^{{-1\over g_s}\,\tr \left[ V_{\vec a_i}(M_i)-V_{\vec a_{i-1}}(M_i) \right]
%+ V_{\vec a_i,\vec a_{i-1}}(X_i) 
} \,\,\,
 \prod_{i=1}^{m} e^{{-1\over g_s}\,\tr \left[V_{\vec a_{i-1}}(M_{i-1})-V_{\vec a_{i}}(M_{i-1}) \right]
%+ \td V_{\vec a_{i-1},\vec a_{i}}(X_{i-1}) 
} \nn \\
&& \prod_{i=1}^{m+1} e^{{1\over g_s} \tr (M_i-M_{i-1})R_i} \,\,\,
 \prod_{i=1}^{m} e^{(S_i+{i\pi\over g_s})\,\tr\, \ln M_i}\,  \nn\\
&& e^{\tr \ln f_{0}(M_0)}\,\,e^{\tr \ln f_{m+1}(M_{m+1})}\,\, \prod_{i=1}^{m} e^{\tr \ln f_{i}(M_i)} \,. \label{m_integral}
\ean
All matrices are taken of size
\beq
N=(n+1)\, d \,,
\eeq
where $d$ is the cut-off discussed in section \ref{secrepresentations}.
We have introduced the notation
\beq
X(\vec \alpha_{m+1})  = {\rm diag} (X(\vec \alpha_{m+1})_i)_{i=1,\dots,N}
\,\, , \qquad
X(\vec \alpha_{m+1})_{j d+k} = q^{h_k(\alpha_{j,m+1})},
\eeq
\beq
X(\vec \alpha_0)  = {\rm diag} (X(\vec \alpha_0)_i)_{i=1,\dots,N}
\,\, , \qquad
X(\vec \alpha_0)_{j d+k} = q^{h_k(\alpha_{j,0})},
\eeq
for $k=1, \ldots, d$, $j=0, \ldots,n$. $\Delta(X)=\prod_{i<j}(X_i-X_j)$ is the Vandermonde determinant. $V_{\vec a_i}(x)$ was introduced in (\ref{partpot}). For $i=1,\dots,m$, we have defined
\beq
f_i(x) =  \prod_{j=0}^n {g(1)^2\,\,e^{({1\over 2}+{i\pi\over \ln q})\, \ln{(x q^{1-a_{j,i}})}}\, \,e^{{(\ln{(x q^{1-a_{j,i}})})^2\over  2 g_s}}\over g(x\,q^{1-a_{j,i}})\, g(q^{a_{j,i}}/x)\,} \,.
\eeq
The denominator of these functions induces simple poles at $x=q^{a_{j,i}+l}$ for $j=0,\dots,n$ and $l\in \mathbb Z$. The numerator is chosen such that they satisfy the relation $f_i(qx)=f_i(x)$. This enforces a simple $l$ dependence of the residues taken at $x=q^{a_{j,i}+l}$, given by a prefactor $q^l$ -- a fact which will be important in the following. These residues are in fact given by
\beqn \label{resf}
 \Res_{q^{a_{j,i}+l}} f_i(x) = q^{a_{j,i}+l}\,\, \hat f_{j,i} =-\, q^{a_{j,i}+l}\,\,   \prod_{k\neq j} {g(1)^2\,\,e^{({1\over 2}+{i\pi\over \ln q})\, (1+a_{j,i}-a_{k,i})\ln q}\, \,e^{{(\ln{(q^{1+a_{j,i}-a_{k,i}})})^2\over  2 g_s}}\over g(q^{a_{j,i}-a_{k,i}})\,(1-q^{a_{k,i}-a_{j,i}}) g(q^{a_{k,i}-a_{j,i}})} \,,
\eeqn
where $\hat f_{j,i}$ is independent of the integer $l$. 

The parameters $S_i$ are defined by 
\beqn \label{si}
S_i =  s_{0,i-1}+t_{0,i-1}= s_{j,i-1} -\sum_{k<j} t_{k,i}+\sum_{k\leq j} t_{k,i-1} \,.
\eeqn
The final equality holds for arbitrary $j$, and can be verified upon invoking (\ref{identify_class}) repeatedly.

For $i=0$ and $i=m+1$, we define
\beq
f_0(x) = {1\over \prod_{j=0}^n \prod_{i=1}^d (x-q^{h_i(\alpha_{j,0})})} \,,
\eeq
\beq
f_{m+1}(x) = {1\over \prod_{j=0}^n \prod_{i=1}^d  (x-q^{h_i(\alpha_{j,m+1})})} \,.
\eeq
Notice that if the representations $\vec\alpha_0$ or $\vec\alpha_{m+1}$ are trivial, i.e. $h_i(\alpha_{j,0})=d-i+a_{j,0}$ or $h_i(\alpha_{j,m+1})=d-i+a_{j,m+1}$, we have
\beq
f_0(x) = \prod_{j=0}^n {g(x\,q^{1-a_{j,0}-d})\over x^d\, g(x\,q^{1-a_{j,0}}) }\,,\hspace{1cm} f_{m+1}(x) = \prod_{j=0}^n {g(x\,q^{1-a_{j,m+1}-d})\over x^d\, g(x\,q^{1-a_{j,m+1}}) }
\eeq
respectively.
The functions $f_0$ and $f_{m+1}$ have simple poles
at $x=q^{h_l(\alpha_{j,0})}$ (resp. $x=q^{h_l(\alpha_{j,m+1})}$) for $l=1,\dots,d$, with residue
\beq
\hat f_{j,0;l} = \Res_{q^{h_l(\alpha_{j,0})}} f_0(x) =   {1\over \prod_{j'\neq j} \prod_{i=1}^d (q^{h_l(\alpha_{j,0})}-q^{h_i(\alpha_{j',0})})} \,{1\over  \prod_{i\neq l} (q^{h_l(\alpha_{j,0})}-q^{h_i(\alpha_{j,0})})} \,,
\eeq
\beq
\hat f_{j,m+1;l} = \Res_{q^{h_l(\alpha_{j,m+1})}} f_{m+1}(x) =   {1\over \prod_{j'\neq j} \prod_{i=1}^d (q^{h_l(\alpha_{j,m+1})}-q^{h_i(\alpha_{j',m+1})})} \,{1\over  \prod_{i\neq l} (q^{h_l(\alpha_{j,m+1})}-q^{h_i(\alpha_{j,m+1})})} \,.
\eeq
The $l$ dependence here is more intricate than above, but this will not play any role since the partitions $\alpha_{j,0}$ and $\alpha_{j,m+1}$ are kept fixed, not summed upon.
  
\bigskip

\begin{figure}[h]
 \centering
 \includegraphics[width=8cm]{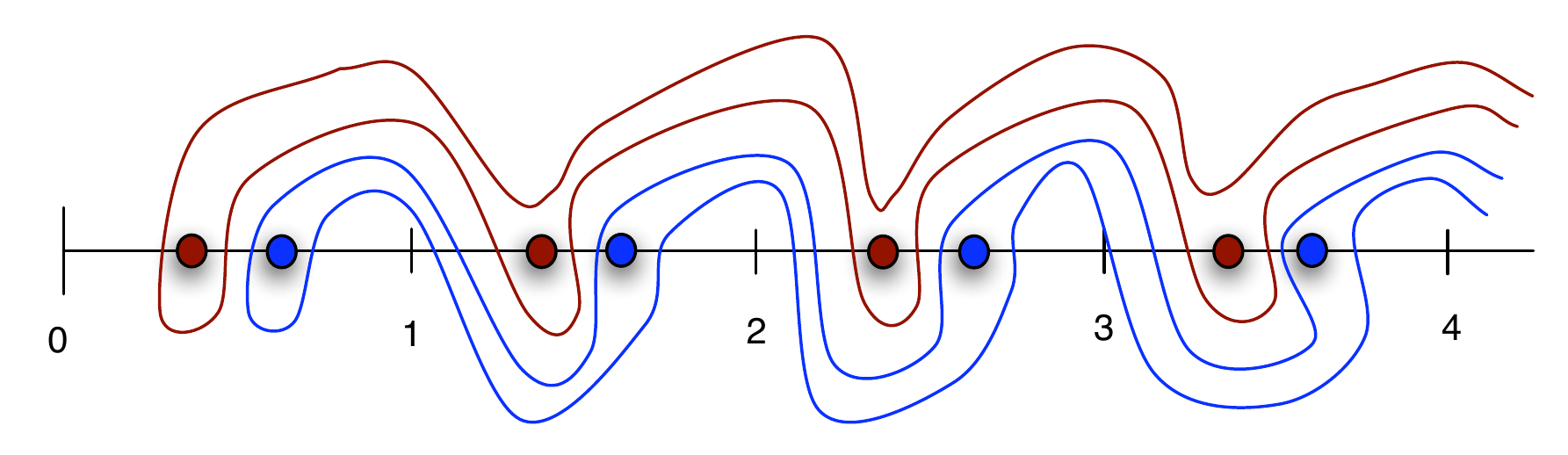}
 \caption{\footnotesize{Two contours surrounding points $a+\mathbb N$ and $b+\mathbb N$, such that  $a-b \notin \mathbb Z$.}}
 \label{contours}
\end{figure}

The integration domains for the matrices $R_i$ are $H_N(\mathbb R_+^N)$, i.e. the set of hermitian matrices  having only positive eigenvalues. For the matrices $M_i, i=1,\dots,m$, the integration domains are $H_N(\Gamma_i)$, where 
\beq
\Gamma_i = \prod_{j=0}^n\, (\gamma_{j,i})^d \,.
\eeq
$\gamma_{j,i}$ is defined as a contour which encloses all points of the form $q^{a_{j,i}+\mathbb N}$, and does not intersect any contours $\gamma_{k,l}$, $(j,i) \neq (k,l)$. For this to be possible, we must require that the differences $a_{j,i}-a_{j',i'}$ be non-integer. The normalized logarithms of two such contours are depicted in figure \ref{contours}.

 We have defined
\beq
H_N(\Gamma_i) = \{ M=U\, \Lambda \, U^\dagger \, , \quad U\in U(N)\, , \,\,\, \Lambda={\rm diag}(\lambda_1,\dots,\lambda_N)\, \in \Gamma_i \} \,,
\eeq
i.e. $H_N(\Gamma_i)$ is the set of normal matrices with eigenvalues on $\Gamma_i$. By definition, the measure on $H_N(\Gamma_i)$ is (see \cite{MehtaBook})
\beqn\label{eqdefdMdUdL}
dM = {1\over N!}\,\, \Delta(\Lambda)^2\,\, dU\,d\Lambda \,,
\eeqn
where $dU$ is the Haar measure on $U(N)$, (normalized not to $1$, but to a value depending only on $N$, such that the Itzykson-Zuber integral evaluates as given in (\ref{eqdefIZ}) with pre\-factor $1$), and $d\Lambda$ is the product of the measures for each eigenvalue along its integration path.

The integration domains for the matrices $M_0$, $M_{m+1}$ are $H_N(\Gamma_0)$, $H_N(\Gamma_{m+1})$ respectively, where 
\beqn  \label{outer_contours}
\Gamma_0 =  (\sum_{j=0}^n \gamma_{j,0})^N \,,
\qquad \quad
\Gamma_{m+1} =  (\sum_{j=0}^n \gamma_{j,m+1})^N \,.
\eeqn

\bigskip

The goal of the rest of this section is to prove that the matrix integral (\ref{m_integral}) reproduces the topological string partition function for target space the fiducial geometry $\CYX_0$.

\subsection{Diagonalization}

Let us first diagonalize all matrices.
We write
\beq
M_i = U_i \, X_i\, U_i^\dagger \,,
\eeq
\beq
R_i = \td U_i \, Y_i\, \td U_i^\dagger,
\eeq
where $U_i$ and $\td U_i$ are unitary matrices.

By the definition (\ref{eqdefdMdUdL}), the measures $dM_i$ and $dR_i$ are given by
\beq
dM_i = {1\over N!}\, \Delta(X_i)^2\, dU_i\, dX_i
\quad , \quad
dR_i = {1\over N!}\, \Delta(Y_i)^2\, d\td U_i\, dY_i  \,.
\eeq

The matrix integral thus becomes
\ba
{\cal Z}_{\rm MM}(Q,g_s,\vec\alpha_{m+1},\vec\alpha_0^T)
&=& {\Delta(X(\vec \alpha_{m+1}))\,\, \Delta(X(\vec \alpha_0))\over (N!)^{2m+3}} \,\, 
\prod_{i=0}^{m+1} \int_{\Gamma_i} dX_i \,\Delta(X_i)^2\,
 \prod_{i=1}^{m+1}\int_{{\mathbb R}_+^N}\,dY_i\,\Delta(Y_i)^2 \cr
&& \prod_{i=0}^{m+1} dU_i\,\,\prod_{i=1}^{m+1} d \td U_i \cr
&& \prod_{i=1}^{m} e^{{-1\over g_s}\,\tr \left[ V_{\vec a_i}(X_i)-V_{\vec a_{i-1}}(X_i) \right]
%+ V_{\vec a_i,\vec a_{i-1}}(X_i) 
} \,\,\,
 \prod_{i=1}^{m} e^{{-1\over g_s}\,\tr \left[V_{\vec a_{i-1}}(X_{i-1})-V_{\vec a_{i}}(X_{i-1}) \right]
%+ \td V_{\vec a_{i-1},\vec a_{i}}(X_{i-1}) 
} \cr
\cr
&& \prod_{i=1}^{m+1} e^{{1\over g_s} \tr X_i U_i^\dagger \td U_i Y_i \td U_i^\dagger U_i}\,\,e^{{-1\over g_s} \tr X_{i-1} U_{i-1}^\dagger \td U_i Y_i \td U_i^\dagger U_{i-1}} 
\,\, \prod_{i=1}^{m} e^{(S_i+{i\pi\over g_s})\,\tr\, \ln X_i}\, \cr
&& e^{\tr \ln f_{0}(X_0)}\,\,e^{\tr \ln f_{m+1}(X_{m+1})}\,\, \prod_{i=1}^{m} e^{\tr \ln f_{i}(X_i)} \,.
\ea
Next, we introduce the matrices $\hat{U_i}$, $\check{U_i}$, for $i=1, \ldots, m+1$, via
\beq
\hat U_i = U_i^\dagger \td U_i 
\quad , \quad
\check U_i = \td U_i^\dagger  U_{i-1} \,.
\eeq
We can express $U_0, \ldots, U_{m+1}$, and $\tilde{U}_1, \ldots, \tilde{U}_{m+1}$, in terms of these matrices and $U_{m+1}$,
\beq
U_i = U_{m+1}\, \hat U_{m+1}\, \check U_{m+1}\,\hat U_{m}\, \check U_{m} \dots \hat U_{i+1}\, \check U_{i+1} \,,
\eeq
\beq
\tilde U_i = U_{m+1}\, \hat U_{m+1}\, \check U_{m+1}\,\hat U_{m}\, \check U_{m} \dots \hat U_{i+1}\, \check U_{i+1}\, \hat U_i \,.
\eeq

With this change of variables, we arrive at
\ba
{\cal Z}_{\rm MM}(Q,g_s,\vec\alpha_{m+1},\vec\alpha_0^T)
&=& {\Delta(X(\vec \alpha_{m+1}))\,\, \Delta(X(\vec \alpha_0))\over (N!)^{2m+3}} \,\, 
\prod_{i=0}^{m+1} \int_{\Gamma_i} dX_i \,\Delta(X_i)^2\,
 \prod_{i=1}^{m+1}\int_{{\mathbb R}_+^N}\,dY_i\,\Delta(Y_i)^2 \cr
&& \int dU_{m+1}\, \prod_{i=1}^{m+1} d\hat U_i\,\,\prod_{i=1}^{m+1} d \check U_i \cr
&& \prod_{i=1}^{m} e^{{-1\over g_s}\,\tr \left[ V_{\vec a_i}(X_i)-V_{\vec a_{i-1}}(X_i) \right]
} \,\,
 \prod_{i=1}^{m} e^{{-1\over g_s}\,\tr \left[V_{\vec a_{i-1}}(X_{i-1})-V_{\vec a_{i}}(X_{i-1}) \right]
} \cr
&& \prod_{i=1}^{m+1} e^{{1\over g_s} \tr X_i \hat U_i Y_i \hat U_i^\dagger }\,\,e^{{-1\over g_s} \tr X_{i-1} \check U_i^\dagger  Y_i \check U_i } 
\,\, \prod_{i=1}^{m} e^{(S_i+{i\pi\over g_s})\,\tr\, \ln X_i}\, \cr
&& e^{\tr \ln f_{0}(X_0)}\,\,e^{\tr \ln f_{m+1}(X_{m+1})}\,\, \prod_{i=1}^{m} e^{\tr \ln f_{i}(X_i)} \,.
\ea

Notice that the integral over $U_{m+1}$ decouples, and $\int dU_{m+1}={\rm Vol}(U(N))$. 

\subsection{Itzykson-Zuber integral and Cauchy determinants}

The $\hat U_i$ and $\check U_i$ appear in the form of Itzykson-Zuber integrals \cite{ItzyksonZuber},
\beqn  \label{eqdefIZ}
I(X,Y) = \int dU\, e^{\tr X U Y U^\dagger} = {\mathop{{\det}}_{p,q} (e^{x_p\,y_q})\over \Delta(X)\,\Delta(Y)} \,,
\eeqn
where $x_p$ and $y_q$ are the eigenvalues of $X$ and $Y$.
We thus have
\ba
{\cal Z}_{\rm MM}(Q,g_s,\vec\alpha_{m+1},\vec\alpha_0^T)
&\propto& {\Delta(X(\vec \alpha_{m+1}))\,\, \Delta(X(\vec \alpha_0))\over (N!)^{2m+3}} \,\, 
\prod_{i=0}^{m+1} \int_{\Gamma_i} dX_i \,\Delta(X_i)^2\,
 \prod_{i=1}^{m+1}\int_{{\mathbb R}_+^N}\,dY_i\,\Delta(Y_i)^2 \cr
&& \prod_{i=1}^{m} e^{{-1\over g_s}\,\tr \left[ V_{\vec a_i}(X_i)-V_{\vec a_{i-1}}(X_i) \right]
} \,\,
 \prod_{i=1}^{m} e^{{-1\over g_s}\,\tr \left[ V_{\vec a_{i-1}}(X_{i-1})-V_{\vec a_{i}}(X_{i-1}) \right]
} \cr
&& \prod_{i=1}^{m+1} I({1\over g_s}X_i,Y_i) \,\, I(-{1\over g_s}X_{i-1},Y_i) 
\,\, \prod_{i=1}^{m} e^{(S_i+{i\pi\over g_s})\,\tr\, \ln X_i}\, \cr
&& e^{\tr \ln f_{0}(X_0)}\,\,e^{\tr \ln f_{m+1}(X_{m+1})}\,\, \prod_{i=1}^{m} e^{\tr \ln f_{i}(X_i)} \cr
&\propto& {\Delta(X(\vec \alpha_{m+1}))\,\, \Delta(X(\vec \alpha_0))\over (N!)^{2m+3}} \,\, 
\prod_{i=0}^{m+1} \int_{\Gamma_i} dX_i \,\,
 \prod_{i=1}^{m+1}\int_{{\mathbb R}_+^N}\,dY_i\, \cr
&& \Delta(X_0)\,\Delta(X_{m+1})\,\,\prod_{i=1}^{m} e^{{-1\over g_s}\,\tr \left[V_{\vec a_i}(X_i)-V_{\vec a_{i-1}}(X_i) \right]
} \cr
&& \prod_{i=1}^{m} e^{{-1\over g_s}\,\tr \left[ V_{\vec a_{i-1}}(X_{i-1})-V_{\vec a_{i}}(X_{i-1}) \right]} \,\,
\,\, \prod_{i=1}^{m} e^{(S_i+{i\pi\over g_s})\,\tr\, \ln X_i}\, 
 \cr
&& \prod_{i=1}^{m+1}  \mathop{{\det}}_{p,q}(e^{{1\over g_s} (X_i)_{p}\, (Y_i)_{q} })
\,\,\, \mathop{{\det}}_{p,q}(e^{{-1\over g_s} (X_{i-1})_{p}\, (Y_i)_{q} })  \cr
&& e^{\tr \ln f_{0}(X_0)}\,\,e^{\tr \ln f_{m+1}(X_{m+1})}\,\, \prod_{i=1}^{m} e^{\tr \ln f_{i}(X_i)} \,,
\ea
where we have dropped an overall sign, powers of $g_s$, and the group volume ${\rm Vol}(U(N))$ which are constant prefactors of no interest to us.

Next, we perform the integrals over $Y_i$ along $\mathbb R_+^N$.
\ba
&& \int_{\mathbb R_+^N} dY \mathop{{\det}}_{p,q}(e^{{1\over g_s} (X_i)_{p}\, (Y)_{q} })
\,\,\, \mathop{{\det}}_{p,q}(e^{{-1\over g_s} (X_{i-1})_{p}\, (Y)_{q} })  \cr
&=& \sum_\sigma \sum_{\td\sigma} (-1)^\sigma (-1)^{\td\sigma}\,\, \prod_{p=1}^N
\int_{0}^\infty dy_p\, e^{{y_p\over g_s}((X_i)_{\sigma(p)}-(X_{i-1})_{\td\sigma(p)})} \cr
&=& \sum_\sigma \sum_{\td\sigma} (-1)^\sigma (-1)^{\td\sigma}\,\, \prod_{p=1}^N {g_s\over (X_{i-1})_{\td\sigma(p)}-(X_i)_{\sigma(p)}} \cr
&=& N!\, g_s^N\,\, \mathop{{\det}}_{p,q}{\left({1\over (X_{i-1})_{p}-(X_i)_{q}}\right)} \,.
\ea
Note that the integral is only convergent for $(X_i)_{\sigma(p)}-(X_{i-1})_{\td\sigma(p)} <0$. For $X_i$ that violate this inequality, we will define the integral via its analytic continuation given in the third line.

An application of the Cauchy determinant formula,
\beq
\det \left( \frac{1}{x_i + y_j} \right)_{1 \le i < j  \le n} = \frac{\prod_{1 \le i < j  \le n} (x_j - x_i) (y_j - y_i)}{\prod_{i,j=1}^n (x_i + y_j)} \,,
\eeq
yields
\beq
\int_{\mathbb R_+^N} dY \mathop{{\det}}_{p,q}(e^{{1\over g_s} (X_i)_{p}\, (Y)_{q} })
\,\,\, \mathop{{\det}}_{p,q}(e^{{-1\over g_s} (X_{i-1})_{p}\, (Y)_{q} })  
= (-1)^{\binom{N}{2}} N!\, g_s^N\,\, {\Delta(X_i)\,\Delta(X_{i-1})\over \Delta(X_{i-1},X_i)} \,,
\eeq
where the notation $\Delta(X_{i-1},X_i)$ was introduced in (\ref{doublevan}). 
Evaluating the $Y_i$ integrals thus, and continuing to drop overall signs and powers of $g_s$, our matrix integral becomes
\ba
{\cal Z}_{\rm MM}(Q,g_s,\vec\alpha_{m+1},\vec\alpha_0^T)
&\propto& {\Delta(X(\vec \alpha_{m+1}))\,\, \Delta(X(\vec \alpha_0))\over (N!)^{m+3}} \,\, 
\prod_{i=0}^{m+1} \int_{\Gamma_i} dX_i \,\,\Delta(X_i)^2 \cr
&& \prod_{i=1}^{m} e^{{-1\over g_s}\,\tr \left[ V_{\vec a_i}(X_i)-V_{\vec a_{i-1}}(X_i) \right]
} \,\,
 \prod_{i=1}^{m} e^{{-1\over g_s}\,\tr \left[V_{\vec a_{i-1}}(X_{i-1})-V_{\vec a_{i}}(X_{i-1}) \right]
} \cr
&& \prod_{i=1}^{m+1}  {1\over \Delta(X_{i-1},X_i)}  \,\, 
\,\, \prod_{i=1}^{m} e^{(S_i+{i\pi\over g_s})\,\tr\, \ln X_i}\, \cr
&& e^{\tr \ln f_{0}(X_0)}\,\,e^{\tr \ln f_{m+1}(X_{m+1})}\,\, \prod_{i=1}^{m} e^{\tr \ln f_{i}(X_i)} \,.
\ea

\subsection{Recovering the sum over partitions} \label{recovering_sum}

Following the steps introduced in \cite{KlemmSulkowski} in reverse, we next decompose the diagonal matrix $X_i$ into blocks, $$X_i= {\rm diag}\,(X_{0,i},X_{1,i},\dots,X_{n,i}) \,,$$ where each matrix $X_{j,i}$ is a $d\times d$ diagonal matrix whose eigenvalues are integrated on the contours $\gamma_{j,i}$ surrounding points of the form $q^{a_{j,i}+\mathbb N}$.
We arrive at
\ba
{\cal Z}_{\rm MM}(Q,g_s,\vec\alpha_{m+1},\vec\alpha_0^T)
&\propto& {\Delta(X(\vec \alpha_{m+1}))\,\, \Delta(X(\vec \alpha_0))\over (N!)^{m+3}} \,\, 
\prod_{i=0}^{m+1}\prod_{j=0}^n \int_{(\gamma_{j,i})^d} dX_{j,i} \cr
&& \Delta(X_0)\,\Delta(X_{m+1})\,\, 
\prod_{i=1}^{m+1}  {\Delta(X_{i-1}) \Delta(X_i)\over \Delta(X_{i-1},X_i)}  \cr
&& \prod_{i=1}^{m} e^{{-1\over g_s}\,\tr \left[ V_{\vec a_i}(X_i)-V_{\vec a_{i-1}}(X_i) \right]
} \,\, \prod_{i=1}^{m} e^{{-1\over g_s}\,\tr \left[V_{\vec a_{i-1}}(X_{i-1})-V_{\vec a_{i}}(X_{i-1}) \right]
} \cr
&& e^{\tr \ln f_{0}(X_0)}\,\,e^{\tr \ln f_{m+1}(X_{m+1})}\,\, \prod_{i=1}^{m} e^{\tr \ln f_{i}(X_i)} 
\,\, \prod_{i=1}^{m} e^{(S_i+{i\pi\over g_s})\,\tr\, \ln X_i}\, ,
\ea
with
\beq
 {\Delta(X_{i-1}) \Delta(X_i)\over \Delta(X_{i-1},X_i)}  
= {\prod_{j} \Delta(X_{j,i-1}) \prod_{j} \Delta(X_{j,i}) \, \prod_{j<l} \Delta(X_{j,i-1},X_{l,i-1})\,\,\prod_{j<l} \Delta(X_{j,i},X_{l,i}) \over \prod_{j,l}\Delta(X_{j,i-1},X_{l,i})}  \,.
\eeq

Our next step is to evaluate the $dX_{j,i}$ integrals via Cauchy's residue theorem. The poles of the integrands lie at the poles of $f_i$, and the zeros of $\Delta(X_{i-1}, X_i)$. However, we have been careful to define our contours $\gamma_{j,i}$ in a way that only the poles of $f_i$ contribute. These lie at the points $q^{a_{j,i} + \IN}$. Hence, the integrals evaluate to a sum of residues over the points
\beq
(X_{j,i})_l = q^{a_{j,i}+(h_{j,i})_l} \,,
\eeq
where each $(h_{j,i})_l$ is a positive integer.

Since the integrand contains a Vandermonde of the eigenvalues of $X_{j,i}$, the residues vanish whenever two eigenvalues are at the same pole of $f_i$, i.e. if two $(h_{j,i})_l$ coincide. Moreover, since the integrand is symmetric in the eigenvalues, upon multiplication by $N!$, we can assume that the $(h_{j,i})_l$ are ordered,
\beq
(h_{j,i})_1>(h_{j,i})_2>(h_{j,i})_3> \dots >(h_{j,i})_d \geq 0.
\eeq
The $(h_{j,i})_l$ hence encode a partition $\alpha_{j,i}$ via $(h_{j,i})_l =  (\alpha_{j,i})_l-i+d$, and we have reduced our integrals to a sum over partitions. In terms of the function $h_l(\alpha)$ introduced in (\ref{defh}),
\beq
(X_{j,i})_l = q^{h_l(\alpha_{j,i})}
\,\, , \quad
h_l(\alpha_{j,i}) = (h_{j,i})_l +a_{j,i} \,,
\eeq
\beq
h_1(\alpha_{j,i})>h_2(\alpha_{j,i})>\dots>h_d(\alpha_{j,i})\geq a_{j,i}.
\eeq

\medskip
Notice that unlike $f_i$, $i=1, \ldots, m$, $f_0$ and $f_{m+1}$ only have a finite number of $N=(n+1)d$ poles. Since the $(h_{j,0})_l$, $(h_{j,m+1})_l$ respectively can be chosen pairwise distinct and ordered, $f_0$ and $f_{m+1}$ act as delta functions in the integrals over the $N\times N$ matrices $X_0$ and $X_{m+1}$, and fix these to the prescribed values $X(\vec\alpha_0)$ and $X(\vec\alpha_{m+1})$ respectively.

\medskip

Performing the integrals hence yields
\ba
{\cal Z}_{\rm MM}(Q,g_s,\vec\alpha_{m+1},\vec\alpha_0^T)
&\propto& \Delta(X(\vec \alpha_{m+1}))^2 \Delta(X(\vec \alpha_0))^2  \\ 
&& \hspace{-2cm}\sum_{\{\alpha_{j,i}|j=0,\dots,n;\, i=1,\dots,m+1\}} \,\,\,  \prod_{i=1}^{m+1}  {\Delta(X(\vec\alpha_{i-1})) \Delta(X(\vec\alpha_i))\over \Delta(X(\vec\alpha_{i-1}),X(\vec\alpha_i))}  \cr
&& \prod_{i=1}^{m} e^{{-1\over g_s}\,\tr \left[V_{\vec a_i}(X(\vec\alpha_i))-V_{\vec a_{i-1}}(X(\vec\alpha_i)) \right]
} \,\, \prod_{i=1}^{m} e^{{-1\over g_s}\,\tr \left[V_{\vec a_{i-1}}(X(\vec\alpha_{i-1}))-V_{\vec a_{i}}(X(\vec\alpha_{i-1})) \right]
} \cr
&& \,\, \prod_{i=1}^{m} e^{(S_i+{i\pi\over g_s})\,\tr\, \ln X(\vec\alpha_i)}  \prod_{i=0}^{m+1}\prod_{j=0}^n\,\prod_{l=1}^d  \,\, \left(\Res_{q^{h_l(\alpha_{j,i})}}\, f_i \right)\, .
\ea
Notice that
\beq
\prod_j \prod_l \Res_{q^{h_l(\alpha_{j,0})}}\, f_0   = {1\over  \Delta(X(\vec \alpha_0))^2} \,,
\eeq
\beq
\prod_j \prod_l \Res_{q^{h_l(\alpha_{j,m+1})}}\, f_{m+1}   = {1\over  \Delta(X(\vec \alpha_{m+1}))^2} \,.
\eeq
Furthermore,
\beq
\Res_{q^{h_l(\alpha_{j,i})}}\, f_i   = q^{h_l(\alpha_{j,i})}\,\,\hat f_{j,i} \,,
\eeq
where $\hat f_{j,i}$ computed in  (\ref{resf}) is independent  of $h_l(\alpha_{j,i})$.  
We thus have
\beq
e^{S_i\, \tr \ln X(\vec \alpha_i)} \,\, \prod_{j=0}^n\,\prod_{l=1}^d  \,\, \left(\Res_{q^{h_l(\alpha_{j,i})}}\, f_i \right) = e^{(S_i+1)\tr \ln X(\vec \alpha_i)}\,\,\prod_{j=0}^n\,(\hat f_{j,i})^d\, .
\eeq
Upon substituting the expression (\ref{si}) for $S_i$, we finally arrive at
\ban
&& {\cal Z}_{\rm MM}(Q,g_s,\vec\alpha_{m+1},\vec\alpha_0^T) \cr
&\propto&  \prod_{i=1}^{m}\prod_{j=0}^n\,(\hat f_{j,i})^d\,\,\,
\sum_{\{\alpha_{j,i}|j=0,\dots,n;\, i=1,\dots,m+1\}} \cr  
&& \prod_{i=1}^{m+1} {\prod_{j} \Delta(X(\alpha_{j,i-1})) \prod_{j} \Delta(X(\alpha_{j,i})) \, \prod_{j<l} \Delta(X(\alpha_{j,i-1}),X(\alpha_{l,i-1}))\,\,\prod_{j<l} \Delta(X(\alpha_{j,i}),X(\alpha_{l,i})) \over \prod_{j,l}\Delta(X(\alpha_{j,i-1}),X(\alpha_{l,i}))}   \cr
&& \prod_{i=1}^{m} e^{{-1\over g_s}\,\tr \left[ V_{\vec a_i}(X(\vec\alpha_i))-V_{\vec a_{i-1}}(X(\vec\alpha_i)) \right]
} \,\,  \prod_{i=1}^{m}\prod_{k=0}^n e^{({1\over 2}-\sum_{j\leq k} (a_{j,i}-a_{j,i-1}) -{i\pi\over g_s} )\tr \ln{X(\alpha_{k,i})} } \cr
&& \prod_{i=1}^{m} e^{{-1\over g_s}\,\tr \left[ V_{\vec a_{i-1}}(X(\vec\alpha_{i-1}))-V_{\vec a_{i}}(X(\vec\alpha_{i-1})) \right]} \,\,  \prod_{i=1}^{m} \prod_{k=0}^n e^{({1\over 2}-\sum_{j< k} (a_{j,i}-a_{j,i+1})  )\tr \ln{X(\alpha_{k,i})} } \cr
&& \,\, \prod_{i=1}^{m}\prod_{j=0}^{n} e^{s_{j,i}\tr\ln X(\alpha_{j,i})}\, .\cr
\ean

Comparing to (\ref{Zstripsum1}) and (\ref{ZvertexprodZstrips}), we conclude
\ba
{\cal Z}_{\rm MM}(Q,g_s,\vec\alpha_{m+1},\vec\alpha_0^T) \propto \sum_{\alpha_{j,i}, j=0,\dots,n;\, i=1,\dots,m}\,\,\,
\prod_{i=1}^{m+1} Z_{\rm strip}(\vec\alpha_i,\vec\alpha_{i-1}^T) \,\, \prod_{j=0}^n\prod_{i=1}^m q^{s_{j,i}\,|\alpha_{j,i}|}  \,, 
\ea
i.e.
\beq
{\cal Z}_{\rm MM}(Q,g_s,\vec\alpha_{m+1},\vec\alpha_0^T) \propto Z_{\rm vertex}({\CYX_0}) = e^{\sum_g g_s^{2g-2} \, GW_g(\CYX_0)} \,.
\eeq

Up to a trivial proportionality constant, we have thus succeeded in rewriting the topological string partition function on the fiducial geometry $\CYX_0$ as a {\em chain of matrices} matrix integral. By our reasoning in section \ref{flops}, this result extends immediately to arbitrary toric Calabi-Yau 3-folds as follows. We have argued that any such 3-fold can be obtained from a sufficiently large choice of fiducial geometry via flops and limits. The respective partition functions are related via (\ref{rel_part_flopped}). Upon the appropriate variable identification, we hence arrive at a matrix model representation of the topological string on an arbitrary toric Calabi-Yau 3-fold.

\section{Implications of our result}  \label{implications}

We have rewritten the topological string partition function as a matrix integral. This allows us to bring the rich theory underlying the structure of matrix models to bear on the study of topological string.

\smallskip
The type of matrix integral we have found to underlie the topological string on toric Calabi-Yau 3-folds is a so-called chain of matrices. This class of models has been studied extensively \cite{Mehta1, MehtaBook}, and many structural results pertaining to it are known.

\subsection{Loop equations and Virasoro constraints}

The loop equations of matrix models provide a set of relations among correlation functions. They are Schwinger-Dyson equations; they follow from the invariance of the matrix integral under a change of integration variables, or by an integration by parts argument.

\medskip

Loop equations for a general chain of matrices have been much studied in the literature, in particular in \cite{David1, DFGZJ,Eynchain, EPrats}. They can be viewed as W-algebra constraints (a generalization of Virasoro constraints) \cite{MiMo}. Having expressed the topological string partition function as a matrix integral, we can hence conclude that Gromov-Witten invariants satisfy W-algebra constraints.

Moreover, a general formal solution of the loop equations for a chain of matrices matrix model was found in \cite{EPrats}, and expressed in terms of so-called symplectic invariants $F_g$ of a spectral curve. The spectral curve for a matrix integral is related to the expectation value of the resolvent of the first matrix in the chain,
\beq
W(x) = \left< \tr {1\over x-M_0}\right>^{(0)} \,.
\eeq
The superscript ${}^{(0)}$ indicates that the expectation value is evaluated to planar order in a Feynman graph expansion. The symplectic invariants $F_g(\curve)$ of an arbitrary spectral curve $\curve$ were defined in \cite{EOFg}. \cite{EPrats} proved that for any chain of matrices integral $Z$, one has
\beq
\ln Z = \sum_g F_g(\curve)
\eeq
with $\curve$ the spectral curve associated to the matrix integral.

\smallskip

Calculating the spectral curve of a chain of matrices matrix model with complicated potentials poses some technical challenges. We will present the spectral curve for our matrix model (\ref{m_integral}) in a forthcoming publication \cite{work_in_progress}.

\subsection{Mirror symmetry and the BKMP conjecture}
The mirror $\hat{{\mathfrak{X}}}$ of a toric Calabi-Yau 3-fold ${\mathfrak{X}}$ is a conic bundle over $\IC^* \times \IC^*$. The fiber is singular over a curve, which we will refer to as the mirror curve ${\cal S}_{\hat {\mathfrak{X}}}$ of $\hat{{\mathfrak{X}}}$. It is a plane curve described by an equation
\beq
{\cal S}_{\hat {\mathfrak{X}}}
 \qquad : \qquad
 H(e^x,e^y)=0 \,,
\eeq
where $H$ is a polynomial whose coefficients follow from the toric data of ${\mathfrak{X}}$ and the K\"ahler parameters of the geometry.

Mirror symmetry is the statement that the topological A-model partition function 
%$$Z_{\mathfrak{X}}(q)=\exp{(\sum_g (\,(\ln{q})^{2g-2} GW_g({\mathfrak{X}}))}$$
with target space ${\mathfrak{X}}$ is equal to the topological B-model partition function with target space $\hat{{\mathfrak{X}}}$.

\smallskip

Extending work of Mari\~no \cite{marino2} proposing a relation between the formalism of \cite{EOFg} and open and closed topological string amplitudes, Bouchard, Klemm, Mari\~no and Pasquetti (BKMP) conjecture in \cite{BKMP} that
\beq
%\encadremath{
GW_g({\mathfrak{X}}) \stackrel{?}{=} F_g({\cal S}_{\hat{{\mathfrak{X}}}}) \,.
\eeq
Here, the $F_g$'s are the symplectic invariants introduced in \cite{EOFg}.
The main interest of this conjecture is that it provides a systematic method for computing the topological string partition function, genus by genus, away from the large radius limit, and without having to solve differential equations.

\smallskip

This conjecture was motivated by the fact that symplectic invariants have many intriguing properties reminiscent of the topological string free energies. They are invariant under transformations ${\cal S}\to \td{\cal S}$ which conserve the symplectic form $dx\wedge dy = d\td{x} \wedge d\td{y}$, whence their name \cite{EOFg}. They satisfy holomorphic anomaly equations \cite{eynhaeq}, they have an integrable structure similar to Givental's formulae \cite{GiventalSemisimple, GiventalHierarchies, AMM1, AMM2, Orantin}, they satisfy some special geometry relations, WDVV relations \cite{CMMV}, and they give the Witten-Kontsevich theory as a special case \cite{EOFg, eynMgnkappa}.

\medskip

BKMP succesfully checked their claim for various examples to low genus.

The conjecture was proved for arbitrary genus in \cite{eynLP} for ${\mathfrak{X}}$ a Hirzebruch rank 2 bundle over $\mathbb P^1$ (this includes the conifold).  Marshakov and Nekrasov \cite{MarshakovNekrasov} proved $F_0=GW_0$ for the family of $SU(n)$ Seiberg-Witten models.  Klemm and Su\l kowski \cite{KlemmSulkowski}, generalizing \cite{eynLP} to Nekrasov's sums over partitions for $SU(n)$ Seiberg-Witten gauge theories, proved the relation for $F_0$, building on work in \cite{NekrasovOkounkov}. In fact, it appears straightforward to extend their computation to arbitrary genus $F_g$. In \cite{Sulkowski}, Su\l kowski provided a matrix model realization of $SU(n)$ gauge theory with a massive adjoint hypermultiplet, again using a generalization of \cite{eynLP} for more general sums over partitions. Bouchard and Mari\~no \cite{BouchardMarino} noticed that an infinite framing limit of the BKMP conjecture for the framed vertex $\CYX=\mathbb C^3$ implies another conjecture for the computation of Hurwitz numbers, namely that the Hurwitz numbers of genus $g$ are the symplectic invariants of genus $g$ for the Lambert spectral curve $e^x=y\, e^{-y}$. That 
 conjecture was proved recently by another generalization of \cite{eynLP} using a matrix model for summing over partitions \cite{Borot:2009ix}, and also by a direct cut and join combinatorial method \cite{EMS}.
The BKMP conjecture was also proved for the framed vertex $\CYX=\mathbb C^3$ in \cite{Zhou, Lin}, using the ELSV formula and a cut and join combinatorial approach.

\medskip

Since we have demonstrated that the topological string partition function is reproduced by a matrix model, we can conclude that the Gromov-Witten invariants coincide with the symplectic invariants 
\beq
\sum_g g_s^{2g-2}\,GW_g = \sum_G F_g(\curve) \,,
\eeq
with $\curve$ the spectral curve of our matrix model. We will compute $\curve$ explicitly in a forthcoming work \cite{work_in_progress}, and demonstrate that it indeed coincides, up to symplectic transformations, with the mirror curve ${\cal S}_{\hat{{\mathfrak{X}}}}$, thus proving the BKMP conjecture for arbitrary toric Calabi-Yau 3-folds, in the large radius limit.

\subsection{Simplifying the matrix model} \label{simplifying_potential}
The matrix models associated to the conifold or to geometries underlying Seiberg-Witten theory have a remarkable property: the spectral curve is the same (perturbatively and up to symplectic transformations) as the one of a simpler matrix model with all $g$-functions replaced by only the leading term in their small $\ln q$ expansion. We will demonstrate in a forthcoming work \cite{work_in_progress} that this property also holds for our matrix integral (\ref{m_integral}). We  can hence simplify the potentials of our matrix model, arriving at
\ba
{\cal Z}_{\rm simp}(Q,g_s,\vec\alpha_{m+1},\vec\alpha_0^T)
&=& \Delta(X(\vec \alpha_{m+1}))\,\, \Delta(X(\vec \alpha_0)) \,\, 
\prod_{i=0}^{m+1} \int_{H_{\bar n_i}(\Gamma_i)} dM_i \,
 \prod_{i=1}^{m+1}\int_{H_{\bar n}({\mathbb R}_+)}\,dR_i \cr
&& \prod_{i=1}^{m} e^{{1\over g_s}\, \tr \sum_{j=0}^n (Li_2(q^{a_{j,i}}/M_i)-Li_2(q^{a_{j,i-1}}/M_{i})) } \cr
&& \,\,\, \prod_{i=0}^{m-1} e^{{1\over g_s}\, \tr \sum_{j=0}^n (Li_2(q^{a_{j,i}}/M_i)-Li_2(q^{a_{j,i+1}}/M_{i})) } \cr
&& \prod_{i=1}^{m+1} e^{{1\over g_s} \tr (M_i-M_{i-1})R_i} \,\,\,
 \prod_{i=1}^{m} e^{(S_i+{i\pi\over g_s})\,\tr\, \ln M_i}\,,
\ea
where the matrix $M_i$ is of size $\bar n_i =\sum_j \bar n_{j,i}$.
\medskip

\subsubsection*{Classical limit}

In the classical limit, the dilogarithm $\Li_2$ becomes the function $x\ln{x}$, and we have
\ba
{\cal Z}_{\rm eff.\, cl}(Q,g_s,\vec\alpha_{m+1},\vec\alpha_0^T)
&=& \Delta(X(\vec \alpha_{m+1}))\,\, \Delta(X(\vec \alpha_0)) \,\, 
\prod_{i=0}^{m+1} \int_{H_{\bar n_i}(\Gamma_i)} dM_i \,
 \prod_{i=1}^{m+1}\int_{H_{\bar n}({\mathbb R}_+)}\,dR_i \cr
&& \prod_{i=1}^{m} e^{{1\over g_s}\, \tr \sum_{j=0}^n (M_i-a_{j,i})\,\ln{(a_{j,i}-M_i)} - (M_i-a_{j,i-1})\,\ln{(a_{j,i-1}-M_i)}
 } \cr
&& \prod_{i=0}^{m-1} e^{{1\over g_s}\, \tr \sum_{j=0}^n (M_i-a_{j,i})\,\ln{(a_{j,i}-M_i)} - (M_i-a_{j,i+1})\,\ln{(a_{j,i+1}-M_i)}
 } \cr
&& \prod_{i=1}^{m+1} e^{{1\over g_s} \tr (M_i-M_{i-1})R_i} \,\,\,
 \prod_{i=1}^{m} e^{(S_i+{i\pi\over g_s})\,\tr\, \ln M_i}\, .
\ea

This model shares features with the Eguchi-Yang matrix model \cite{EguchiYang}, see also \cite{MarshakovNekrasov}.

%, and is often presented as a candidate for an effective matrix model for gauge theories.

\section{Conclusion}  \label{conclusions}

We have rewritten the topological vertex formula for the partition function of the topological A-model as a matrix integral.

Having expressed the topological string in terms of a matrix model, we can bring the immense matrix model toolkit which has been developed since the introduction of random matrices by Wigner in 1951 to bear on questions concerning the topological string and Gromov-Witten invariants. We already started down this path in section \ref{implications} above. Going further, we can apply the method of bi-orthogonal polynomials \cite{MehtaBook} to our matrix model to unearth the integrable system structure (Miwa-Jimbo \cite{MiwaJimbo1, MiwaJimbo2}) underlying the topological string, at least in the case of toric targets, together with its Lax pair, its Hirota equations (which arise as orthogonality relations), etc. In a related vein, free fermions \cite{Harnad-Orlov2, Kostov2} arise in the theory of matrix models when invoking determinantal formulae to express the matrix model measure \cite{EynMehta}.  It will be very interesting to explore how this is related to the occurrence of free fermions in topological string theory, as studied in \cite{ADKMV, KashaniPoor, DHSV, DHS}. More generally, one should study what can be learned about the non-perturbative topological string from its perturbative reformulation as a matrix model, as in the works \cite{MarinoSchiappaWeiss, Marino, EynardMarino,KlemmMarinoRauch}. A recurrent such question, which could be addressed in the matrix model framework (in fact, it was already latently present in the calculations in this work), is that of the quantization of K\"ahler parameters.

On a different note, notice that the matrix model derived in this article, with a potential which is a sum of logs of $q$-deformed $\Gamma$ functions, looks very similar to the matrix model counting plane partitions introduced in \cite{EynPP}. This is a hint that it could be possible to recover the topological vertex formula, corresponding to the topological string with target $\IC^3$ and appropriate boundary conditions, directly from the matrix model approach. Either along these lines or the lines pursued in this paper, it would be interesting to derive a matrix model related to the Nekrasov deformation \cite{Nekrasov,IqbalKozcazVafa} of the topological string.

A completely open question is whether the close relation between topological strings and matrix models persists beyond toric target spaces, and more ambitiously yet, whether there exists a general notion of geometry underlying matrix models.

\section*{Acknowledgments}
B.E. and O.M. would like to thank M. Bertola, J. Harnad, V. Bouchard,  M. Mari\~ no, M. Mulase,  H.~ Ooguri, N.~Orantin, B. Safnuk, for useful and fruitful discussions on this subject. A.K. would like to thank V.~Bouchard and I.~Melnikov for helpful conversations. The work of B.E. is partly supported by the Enigma European network MRT-CT-2004-5652, ANR project GranMa "Grandes Matrices Al\'eatoires" ANR-08-BLAN-0311-01,  
by the European Science Foundation through the Misgam program,
by the Quebec government with the FQRNT. 
B.E. would like to thank the AIM, as well as the organizers and all participants to the workshop \cite{AIM} held at the AIM june 2009.
O.M. would like to thank the CRM (Centre de recheche math\'ematiques de Montr\'eal, QC, Canada) for its hospitality.

\bigskip

%\vfill
%\eject
\appendix{}

\section{q-product}
\label{appgq}

The $g$-function, which plays a central role in the definition of our matrix model, is defined as an infinite product,
\beq
g(x) = \prod_{n=1}^\infty (1-{1\over x}\, q^n) \,.
\eeq
It is the quantum Pochhammer symbol $g(x)= [q/x;q]_{\infty}$, and it is related to the $q$-deformed gamma function via $\Gamma_q(x) = (1-q)^{1-x}\,g(1)/g(q^{1-x})$.

The RHS is convergent for $|q|<1$ and arbitrary complex $x \neq 0$. $g(x)$ satisfies the functional equation
\beq
g(qx)=(1-{1\over x})\,g(x) \,.
\eeq
For $n\in \IN$, we have
\beq
g(q^n) = 0
\eeq and
\beq
g'(q^n) = (-1)^{n-1} g(1)\,q^{-{n(n+1)\over 4}}\,\, \prod_{m=1}^{n-1} [m]
 = g(1)\,q^{-{n(n+1)\over 2}}\,\, [n-1]!
= (-1)^{n-1}\,q^{-{n(n+1)\over 2}}\, {g(1)^2\over g(q^{1-n})} \,.
\eeq

 % This function can also be defined in terms of the so-called quantum $\Gamma$-function $\Gamma_q$,
% \beq
% g(x) = {q^{-\frac{1}{24}} \eta(q) \,\,e^{-{1\over 4\ln q}\ln x \, \ln{qx}}\over (1-{1\over x})\,\Gamma_q({1\over x})}.
% \eeq
% $\Gamma_q$ is a deformation of the ordinary $\Gamma$-function. For integer $n$, it satisfies the property
% \beq
% \Gamma_q(q^{n+1}) = \prod_{j=1}^n [j]  = [n]! \,,
% \eeq
% where the $q-$numbers $[x]$ is defined by
% \beq
% [x] = q^{-x\over 2}-q^{x\over 2} \,.
% %\over q^{-1\over 2}-q^{1\over 2}}
% \eeq
% Ordinary functions can be recovered from their $q$-deformation in the $q \rightarrow 1$ limit. The quantum $\Gamma$-function thus satisfies the functional equation
% \beq
% \Gamma_q(q^{h+1}) = [h]\,\Gamma_q(q^h) \,,
% \eeq
% the $q$-deformation of the relation $\Gamma(h+1)=h\Gamma(h)$ which is recovered at $q=1$. 

Via the triple product representation of the theta function,
\beq
\theta(z;\tau) = \prod_{m=1}^\infty ( 1 - e^{2 \pi i m \tau}) (1 + e^{(2m-1)\pi i \tau + 2 \pi i z})(1 + e^{(2m-1)\pi i \tau - 2 \pi i z})
\,,
\eeq
we obtain the identity
\beq
\theta \left(\frac{1}{2} + \frac{1}{4\pi i} \ln \frac{q}{x^2}\, ;\,\, \frac{\ln q}{2\pi i} \right) = g(x) g(\frac{q}{x}) g(1) \,.
\eeq

We have
\beq
{g(x) \, g(q/x)\over g(1)^2\,\,\sqrt x}\,\,e^{(\ln{x})^2\over 2\ln q}\,\, e^{-i\pi \ln x\over \ln q}   = {-\ln q\over \theta'({1\over 2} - {i\pi\over \ln q}, -{2i\pi\over \ln q}) }\,\,\, \theta\left({\ln x\over \ln q} +{1\over 2} - {i\pi \over \ln q}, {-2i\pi\over \ln q}\right)\,\, 
\eeq
% \beq
% {e^{-i\pi\ln x\over \ln q}\over \Gamma_q(q/x)\,\,\Gamma_q(x)} = {-\ln q\over \theta'({1\over 2} - {i\pi\over \ln q}, -{2i\pi\over \ln q}) }\,\,\, \theta\left({\ln x\over \ln q} +{1\over 2} - {i\pi \over \ln q}, {-2i\pi\over \ln q}\right)\,\, 
% \eeq
where $\theta$ is the Riemann theta-function for the torus of modulus $-2i\pi/\ln q$.
% This relationship is the quantum deformation of $e^{-i\pi x}/\Gamma(1-x)\Gamma(x) = \sin{(\pi x)}/\pi$ for the classical Gamma-function.

At small $\ln q$, the following expansion is valid,
\beq
\ln{g(x)} = {1\over \ln q}\,\sum_{n=0}^\infty {(-1)^n\,B_n\over n!}\,(\ln q)^n\,\, \Li_{2-n}(1/x) \,,
\eeq
where we have used the definition of the Bernoulli numbers $B_n$ as the coefficients in the expansion of $t/(e^t-1)$,
\beq
\frac{t}{e^t -1} = \sum_{n=0}^\infty B_n \frac{t^n}{n!} \,.
\eeq
$\Li_n$ is the polylogarithm function, defined as
\beq
\Li_n(x) = \sum_{k=1}^\infty\, {x^k\over k^n} \,.
\eeq
This is a generalization of the logarithm function, recovered at $n=1$,
\beq
\Li_1(x) = -\ln{(1-x)} \,.
\eeq
It satisfies the functional relation
\beqn
\Li'_n(x) = \frac{1}{x} \Li_{n-1}(x) \,.  \label{derpoly}
\eeqn
Note in particular that this implies that $\Li_n$ is an algebraic function of $x$ for $n \le 0$. E.g.,
\beq
\Li_0(x) = {x\over 1-x} \,.
\eeq
We also define the function
\beq
\psi_q(x) = x\,{g'(x)\over g(x)}  \,.
\eeq
Using the functional equation (\ref{derpoly}) of the polylogarithm, we find its small $\ln(q)$ expansion
\ba
\psi_q(x) &=& -\,{1\over \ln q}\,\sum_{n=0}^\infty {(-1)^n\,B_n\over n!}\,(\ln q)^n\,\, \Li_{1-n}(1/x) \\
&=&  {1\over \ln{q}}\,\Big[
\ln{(1-{1\over x})} -{\ln q\over 2(x-1)}
-\sum_{n=1}^\infty {B_{2n}\over (2n)!}\,\, (\ln q)^{2n}\,\Li_{1-2n}(x) \Big] \,.
\ea
For the second equality, we have used $B_0=1, B_1 = -\frac{1}{2}$, and $B_{2n+1} = 0$ for $n >1$.

\medskip
We have near $x\to\infty$
\beq
\psi_q(x) \sim {q\over 1-q}\,\,{1\over x} + O(x^{-2})
\eeq
and near $x\to 0$:
\beq
\psi(x) \sim {1\over 2}+{i\pi+\ln x\over g_s} +O(x).
\eeq

\bibliography{bibl}
\bibliographystyle{utcaps}

\end{document}